\documentclass[11pt,a4paper]{article}
\usepackage{longtable}
\usepackage{amsmath}
\usepackage{amssymb}
\usepackage{graphicx}
\usepackage{longtable}
\usepackage[labelsep=endash]{caption}
\usepackage{float}
\usepackage[square,comma,sort&compress,numbers]{natbib}
\usepackage{bm}

\newcommand*{\bea}{\begin{eqnarray}}
\newcommand*{\eea}{\end{eqnarray}}
\newcommand*{\be}{\begin{equation}}
\newcommand*{\ee}{\end{equation}}

\newcommand{\bma}{\begin{pmatrix}}
\newcommand{\ema}{\end{pmatrix}}
\usepackage[square,comma,sort&compress,numbers]{natbib}
\title{Two and three point functions in real singlet and complex doublet scalar model}
\author{Muhammad Saad \footnote{muhammadsaad937@gmail.com}, Tajdar Mufti \footnote{tajdar.mufti@gmail.com, tajdar.mufti@lums.edu.pk} \\ Lahore University of Management Sciences\\ Opposite Sector U, D.H.A, Lahore Cantt., 54792, Pakistan}
\begin{document}
\maketitle
\begin{abstract}
We study dynamics of an interacting theory of a real singlet scalar and an $SU(2)$ symmetry preserving complex doublet scalar fields using lattice simulations over a broad  region of the parameter space. The model contains all $SU(2)$-preserving quartic and a real singlet-$SU(2)$ invariant doublet field Yukawa interactions. The field propagators and the Yukawa vertex are the central structures for the study. We complement the study by implementing machine learning routines to the correlation functions in order to probe the underlying physics. The model reveals the role of nonperturbative effects in terms of infrared enhanced field propagators with a stable amputated Yukawa vertex, and a continuum-like scaling window in the parameter space with large regulator. The ultraviolet effects are visible in terms of an scaling structure appearing in a composite operator of mixed scalar interactions and the singlet expectation value.
\end{abstract}
\section{Introduction} \label{section:Introduction}
The scalar sector is an important cornerstone of quantum field theory from the perspectives of both phenomenological searches for new physics and for advancing theoretical understanding of field theories. The discovery of the Higgs boson \cite{Aad:2012tfa,Chatrchyan:2012xdj} underscores the importance of a thorough understanding of the role of scalar interactions within the sector and beyond in order to study extensions beyond the Standard Model (SM) \cite{ATLAS:2023tkl,Cirigliano:2013lpa,Adhikari:2020vqo,Guo:2010hq}. The existence of a real singlet scalar field is the simplest possibility for new physics, and a three point interaction involving one real singlet scalar is the simplest portal to couple the field to other fields.
\par
Field theory models containing scalar interactions have already been studied for various reasons \cite{Lawrence:2022afv,Ye:2025zhs,Habibolahi:2022rcd,deLima:2024lfc,Ghorbani:2018yfr,Wu:2016mbe} which include understanding the phase transition \cite{De:2005ny,Zarikas:1995qb,Jersak:1985nf}, dark matter related physics \cite{Kim:2023pwf,Mondino:2020lsc,Ahmed:2022tfm,Ahmed:2022qeh,Yu:2024xsy,Profumo:2010kp}, the Higgs portal \cite{Lebedev:2021xey,Arcadi:2021mag,Kim:2023bbs,Djouadi:2012zc}, and the electroweak sector \cite{Maas:2013aia,Maas:2014pba} in the SM. In most cases, however, either smaller models are studied which may pertain to a larger scalar sector, or the study is based on perturbation theory \cite{Boehm:2020wbt} to address phenomenology related aspects at weak coupling strength. As presence of other interactions at different coupling strengths, and the corresponding symmetries may significantly effect the overall behavior of the model, and likely the relevant observables, it demands a detailed study of the sub-theories as well as the larger (parent) theory in different regions of the respective parameter space. It requires nonperturbative methods to include strong coupling regime, a task which perturbation theory \cite{Schwartz:2013pla,Kaku:1993ym} can not accomplish. Though, such attempts have also been made, see \cite{vandeBruck:2022xbk,Lewis:2024yvj,Duerr:2015aka,Barger:2007im,Niemi:2024axp} for instance, understanding of the scalar sector remains far from conclusive. An important example is the role of scalar Yukawa interaction in the dynamics of scalar sector and the mathematical structure it conforms to in the corresponding parameter space.
\par
For nonperturbative methods, lattice simulation (LS) \cite{Ruthe:2008rut,Gattringer:2010gl} and the method of Dyson Schwinger equations (DSEs) \cite{Roberts:1994dr} are two well known approaches. They can be viewed as complementing each other as relevant results from lattice simulations may be used in DSEs as known quantities. The field propagators are among strong candidates for this purpose as they can be computed with relative ease and decent precision. Furthermore, they contain crucial information such as capacity of generating dynamical mass \cite{Cornwall:1981zr,Treml:1990cc}, effects due to renormalization scheme, confinement \cite{Chaichian:2018cyv}, and relevance of perturbation theory in general. If computed at large number of points in the parameter space, machine learning \cite{MarslandML} becomes a highly interesting tool to extract the corresponding representative functions. It offers a glimpse into infrared region \footnote{For the sake of clarity, infrared region is taken as the lowest accessible momentum region rather than the extremely small momentum region.} which may not be easily accessible in lattice simulation of moderate lattice size. The region is particularly informative in investigations pertaining to Quantum Chromodynamics (QCD) \cite{Oliveira:2007dy,Cucchieri:2007zm,Fister:2013bh} physics which involve cubic and quartic interactions. Hence, a model containing cubic and quartic interactions invites studying the region.
\par
In this article, we consider a theory of real scalar singlet and an $SU(2)$ preserving complex doublet, termed as the Higgs boson for convenience, containing quartic interactions and an $SU(2)$ preserving Yukawa vertex. It provides a useful framework to examine how a theory containing a trivial subsector is modified in the presence of Yukawa interaction within the scalar sector. We adopted the method of lattice simulations for the study. Our analysis focuses primarily on the field propagators and the Yukawa vertex, with two central objectives: first, to extract representative functional forms for these correlation functions, and second, to complement the lattice simulation results in order to probe the underlying dynamics across the model parameter space. Particular emphasis is placed on the low (lattice) mass behavior of the singlet scalar and Higgs propagators in search of mass-like contributions, the impact of Yukawa interactions, associated with a distinct symmetry structure, and the role of the singlet scalar quartic self-interaction coupling in shaping the dynamics of the theory. Additional correlation functions and field expectation values are also examined to further elucidate the details in the model. Understanding mass-like contributions using both LS and ML propagators, role of Yukawa interaction bearing a different symmetry, and the role of the singlet scalar quartic self coupling in the model constitute our line of investigation. In order to extract the representative functional forms, we employ supervised machine learning (SML) on the statistics collected at $192$ points in the parameter space at different strengths of couplings and scalar (lattice) mass.
\par
The diversity of interactions in the model invites a thorough study of several other aspects including spectroscopy in the $0^{+}$ channel and triviality. We postpone addressing them at this point, and proceed with an assumption that the model is not trivial. Furthermore, during the entire study the cubic self interaction structure of the singlet scalar is ignored.
\section{Technical Details} \label{section:TechnicalDetails}
\begin{table}[ht]
 \caption{Masses and couplings in lattice units for the study.}
 \centering
 \begin{tabular}{| c | c | c | c |}
  \hline
  $m_{s}$ & $\lambda$ & $\alpha$ & $\gamma$ \\
  \hline
  10.0 & 1.0 & 1.0 & 1.0 \\
  1.0 & 0.6 & 0.6 & 0.1 \\
  $10^{-3}$ & 0.001 & $10^{-6}$ & 0.01 \\
  $10^{-6}$ &  & $10^{-12}$ & 0.0 \\
\hline
 \end{tabular}\label{table:masses-n-couplings}
\end{table}
The action in continuum (Minkowski space) is given by
\begin{equation}
\begin{split}
S_{M} = \int d^{4}x [\ \frac{1}{2} \eta^{\mu \nu} \partial_{\mu} \phi \partial_{\nu} \phi  - \frac{1}{2} m^{2}_{s} \phi^{2} + \eta^{\mu \nu} \partial_{\mu} h^{\dagger} \partial_{\nu} h  -  m^{2}_{h} h^{\dagger} h \\ -\lambda \phi h^{\dagger} h -\alpha \phi^{4} - \beta (h^{\dagger}h)^{2} - \gamma \phi^{2} h^{\dagger} h  ]\
\end{split}
\label{eq:CAction}
\end{equation}
The study assumes no contribution from cubic self interactions of the real singlet field. The corresponding unimproved lattice action \footnote{The notations for bare masses and parameters in the continuum and the lattice version of actions are the same. From now on, the masses and parameters are to be taken in lattice version unless otherwise stated.} \cite{Ruthe:2008rut} in Euclidean space is given by
\begin{equation} 
\begin{split}
S= \sum_{n} [\ (8+m^{2}_{h}) h^{*i}(n)h^{i}(n) - 2 \sum_{\mu, i} \mathcal{R} (\ h^{*i}(n+\hat{\mu})h^{i}(n) )\ + \\ \frac{1}{2}(8+m^{2}_{s}) \phi^{2}(n) - 2 \sum_{\mu} \phi(n+\hat{\mu})\phi(n) + \lambda  \phi(n)\sum_{i}h^{*i}(n)h^{i}(n) + \alpha \phi^{4}(n) + \\ \beta (\sum_{i} h^{*i}(n)h^{i}(n))^{2} + \gamma  \phi^{2}(n) \sum_{i}h^{*i}(n)h^{i}(n)  ]\ .
\end{split}
\label{eq:LAction}
\end{equation}
The Higgs field retains SU(2) symmetry while the scalar singlet has trivial symmetry in the model. For the case $\lambda \neq 0$, the presence of scalar Yukawa interaction spoils the symmetry under $\phi(x) \rightarrow -\phi(x)$. Hence, the model is studied with different values of $\lambda$, $\alpha$, $\gamma$, and $m_{s}$ while excluding the region with $\lambda = 0$. Furthermore, $m_{h}$ and $\beta$ are kept fixed at $125.09$ and $0.51639$, respectively, and the region $m_{s} < m_{h}$ is aimed for the study.
In total, $192$ points in the parameter space are considered, the points were chosen from the parameter values given on Table \ref{table:masses-n-couplings} while the simulations were performed on the symmetric lattice of size $18^{4}$ \footnote{At this lattice size, acceptable stability in lattice results is observed in comparison to the size $24^{4}$ lattice. Hence, the chosen size is suitable for both machine learning and vertex computation in entire plane.}. The upper limit of each coupling is fixed at $1.0$ magnitude which is generally the breaking point of perturbation theory. However, the choice of the lower limit is such that the lowest value of the couplings is at least in harmony with perturbative picture. As mentioned earlier, it is the Yukawa interaction which we are closely monitoring, we have selected a $\gamma=0$ but not $\alpha=0$ because we also intend to retain the sub-trivial theory in the model. However, a significantly low value of $\gamma$ is included to observe the region of the parameter space in which the trivial sub-theory has significantly low effect which is expected to appear in lattice spacing (regulator).
\par
The presence of Yukawa interaction in the model tends to raise concern regarding the action changing the sign. To further clarify it, the potential term of the action in continuum can be rewritten as 
\begin{equation}
\begin{split}
\int d^{4}x [\ \lambda \phi h^{\dagger} h + \beta (h^{\dagger}h)^{2} + \frac{1}{2} m^{2}_{s} \phi^{2} +  m^{2}_{h} h^{\dagger} h + \gamma \phi^{2} h^{\dagger} h ]\ \\ = \int d^{4}x [\ \beta (h^{\dagger}h + \frac{\lambda \phi}{2 \beta})^{2} + \frac{1}{2} (m^{2}_{s}-\frac{\lambda^{2}}{2 \beta} ) \phi^{2} + m^{2}_{h} h^{\dagger} h  + \gamma \phi^{2}h^{\dagger} h  ]\ 
\end{split}
\label{eq:LActionPotential}
\end{equation}
with the $\lambda$ bounded within unit magnitude and $2 \beta \approx 1$ during the entire investigation, the lowest value of the negative term in the potential term is approximately $-\phi^{2}$. As the histories for the path integral computations are generated by the same random number generator applied on both the Higgs and the scalar singlet fields, a large enough lattice size may have the fields with a wide range of values, depending upon the random number generator and other relevant details. Furthermore, there are three more terms including the one containing $m_{h}$ ($=125$). It indicates that a negative action is highly unlikely in the studied parameter space. In fact, over a billion histories were observed during the study and none of them was found with a negative action. At this point, it is important to note that the vertex has also been considered in other studies, see \cite{Barger:2007im} for example.
\par
\subsection{Lattice simulations}
For the Monte Carlo lattice simulations, each simulation begins with a hot start \cite{Gattringer:2010gl}. The selection performed on the histories is based on multi-hit Metropolis approach while the same update routine is performed on the histories before and after the equilibration point. The routine is as follows: First, one overrelaxation update is applied on the complex Higgs doublet fields $h= \begin{pmatrix} h^{1} \\ h^{2} \end{pmatrix}$ \footnote{$h^{1}$ and $h^{2}$ are the two Higgs flavors.} using a matrix $\kappa M_{o}$ such that $M_{o} \subset SU(2)$ and $\kappa=1$. It is a relatively expensive step, as it is performed during every update routine, but it refrains the Higgs field from unnatural convergence, or numerical entrapment, of their expectation value. It is followed by 5 local updates applied on scalar fields, and then 5 local updates on the Higgs fields. A local update on the Higgs fields is applied using the matrix $\kappa M_{o}$ as described above with $\kappa$ being a non-zero real number which is not fixed at $1$. An advantage is that Higgs fields do not remain confined within a certain radius, i.e. the Higgs fields do not remain compactified. The update sequence results in an average (over the fields and flavors) Pearson correlation coefficient in the vicinity of $0.55$ between two consecutive histories.
\begin{figure} [h]
 \centering
 \includegraphics[width = 1.0\textwidth]{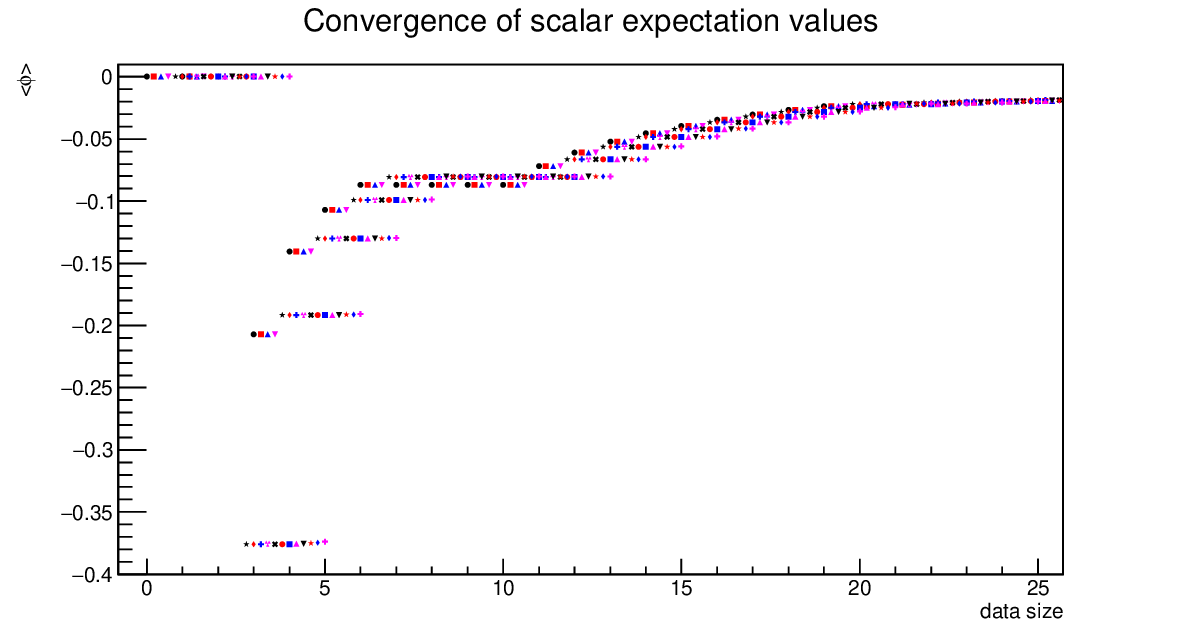}
 \caption{Evolution of scalar expectation values are shown using different hot starts, each represented by a different symbol.}
 \label{fig:equilibration}
\end{figure} 
The equilbration is defined at the completion of $500+20N_{t}$ updates where $N_{t}$ is the lattice size along time dimension (in Euclidean spacetime). The definition is based on convergence of the spatially averaged scalar one point functions, see Figure \ref{fig:equilibration}. In order to maximize memory loss on different hot starts \cite{Gattringer:2010gl}, equilibration point is chosen well ahead of other possibilities in order to improve convergence and stability of the computed quantities \footnote{The ensuing stability supports the definition of equilbration point in terms of other quantities, hence relaxing the stringency of the definition.} Once the equilbration is achieved, measurements of the to-be-determined quantities start. The performed simulations follow the standard lattice simulation algorithm flow available in literature, see \cite{Gattringer:2010gl} for example.
\par
Introducing physical dimensions in the model involves determination of lattice spacing (scale) which is not as straight forward as in other models \cite{Maas:2013aia}. First, the model contains scalar quartic interaction which forms a trivial subtheory \cite{Jora:2015yga}. It means that selecting an operator (directly) depending upon the scalar field may introduce peculiarities in contrast to other choices in the model. Second, there is a possibility of a null ground state in the model which may complicate the situation if the ground state, with $0^{+}$ quantum number, is selected for setting the scale. Third, only the SU(2) preserving field, conveniently termed as the Higgs field in our model, has some relevance to currently known physics. Lastly, higher correlation functions may have larger noise which requires smearing or other methods to reduce the noise. The scale is chosen using a $0^{+}$ operator, which is also gauge invariant, consisting of the Higgs fields. The details are in the following.
\subsubsection{Correlation functions} \label{section:CorrelationFunctions}
A $0^{+}$ operator, which we name the \textit{Higgs ball}, is chosen in order to compute lattice spacing. A operators generally result in a four or higher field correlation function, they require certain measures to reduce statistical fluctuations. In order to achieve it, 2 levels of smearing on every selected history was performed. The following rule was used to smear scalar singlet field at point $\hat{x}^{\mu}$ ($=(\hat{t},\hat{x},\hat{y},\hat{z})$).
\begin{equation}
\begin{split}
\phi_{s+1}(\hat{t},\hat{x},\hat{y},\hat{z}) = \frac{1}{9} [\ \phi_{s}(\hat{t},\hat{x},\hat{y},\hat{z}) + \phi_{s}(\hat{t}+1,\hat{x},\hat{y},\hat{z}) + \phi_{s}(\hat{t}-1,\hat{x},\hat{y},\hat{z})+  \\ \phi_{s}(\hat{t},\hat{x}+1,\hat{y},\hat{z}) + \phi_{s}(\hat{t},\hat{x}-1,\hat{y},\hat{z}) + \phi_{s}(\hat{t},\hat{x},\hat{y}+1,\hat{z}) + \phi_{s}(\hat{t},\hat{x},\hat{y}-1,\hat{z}) + \\ \phi_{s}(\hat{t},\hat{x},\hat{y},\hat{z}+1) + \phi_{s}(\hat{t},\hat{x},\hat{y},\hat{z}-1) ]\
\end{split}
\label{eq:smearing1}
\end{equation}
where $s$ \footnote{Here, $s=0$ represents unsmeared field.} is the level of smearing and $\hat{x}^{\mu} \subset (0,N^{\mu})$. It is assumed that the artifacts on the lattice boundary will diminish for large lattice size. For the study, given our objective of large data collection for machine learning, the highest feasible lattice size is considered for this study. A similar relation was used to perform smearing on each of the two flavors of the Higgs field.
\par
The Higgs ball operator is defined as
\begin{equation}
O^{h}_{s}(t)= \frac{1}{V_{3d}} \int d^{3}x h_{s}^{\dagger}h_{s}.
\label{eq:Higgsballdef1}
\end{equation}
where the spatial dependence is averaged out and the subscript $s$ indicates smearing level (0,1, and 2). Given the observation of 3 exponential terms, combination of 3 exponent terms in $O^{h}_{2}(t)$ were used for curve fitting, and the lowest non-zero mass was chosen to calculate the scale. Mathematically,
\begin{equation}
O^{h}_{2}(t)= c_{1}e^{-m_{1}\hat{t}}+c_{2} e^{-m_{2}\hat{t}}+c_{3} e^{-m_{3}\hat{t}}
\label{eq:Higgsballexpansion1}
\end{equation}
where $m_{1}$, $m_{2}$, $m_{3}$, $c_{1}$, $c_{2}$, and $c_{3}$ are the to-be-determined quantities, and $m_{1} < m_{2} < m_{3}$. Lattice spacing is calculated using the following relation:
\begin{equation}
a=\frac{m_{i}}{M_{h}}
\label{eq:scaleformula}
\end{equation}
where $a$ is the lattice spacing, $M_{h}=125.09$ in $GeVs$, and the $i$th mass is the lowest non-zero lattice mass.
\par
The one point functions (field averaged over spacial directions) and field expectations have certain merits. First, a correlation is suspected between the scalar one point function (or its expectation) and the Yukawa vertex in the model. It may contribute in substantiating the role of the Yukawa vertex. Second, the Higgs expectation value is also found to have connection to the phase structure of a model \cite{Maas:2013aia,Maas:2014pba} which necessitates exploring any relevance in regard to the Yukawa vertex. The one point functions are defined in the following:
\begin{equation}
<\phi (t)> = <\frac{1}{V_{3d}} \int d^{3}x \phi (t,x,y,z) >
\label{eq:spaceavgphi}
\end{equation}
and
\begin{equation}
<h^{i} (t)> = <\frac{1}{V_{3d}} \int d^{3}x h^{i} (t,x,y,z)>
\label{eq:spaceavgH}
\end{equation}
where $i$ is the flavor of the Higgs field and $d^{3}x$ is the integral over the spatial directions. The scalar expectation value is obtained by including the time direction in equation \ref{eq:spaceavgphi}.
\par
The field propagators are renormalized using a scheme used in a previous study \cite{Maas:2013aia}. The scheme for the (flavor diagonal) Higgs propagator is given below:
\begin{equation}
D_{h}(p)= \frac{1}{z(p^{2}+m_{h}^{2})+\delta m^{2}_{h}+\Pi(p)}
\label{eq:spaceavgH1}
\end{equation}
For the two unknown constants, $z$ and $\delta m^{2}$, the following two conditions are used:
\begin{equation}
D_{h}(p) |_{p^{2}=\mu^{2}}= \frac{1}{p^{2}+m^{2}_{h}}  |_{p^{2}=\mu^{2}}
\label{eq:spaceavgH4a}
\end{equation}
\begin{equation}
\frac{\partial}{\partial p} D_{h}(p) |_{p^{2}=\mu^{2}}= \frac{\partial}{\partial p} \frac{1}{p^{2}+m^{2}_{h}} |_{p^{2}=\mu^{2}}
\label{eq:spaceavgH4b}
\end{equation}
where $\mu$ is the renormalization point. The self energy term $\Pi(p)$ is calculated from the unrenormalized propagator $D^{o}_{h}$ using the following equation:
\begin{equation}
D^{o}_{h}(p)= \frac{1}{p^{2}+\Pi(p)}
\label{eq:spaceavgH434}
\end{equation}
The same renormalization scheme is implemented on the scalar propagators, with the replacement of $m_{h}$ by $m_{s}$, $D_{h}$ by $D_{s}$, and $D^{o}_{h}$ by $D^{o}_{s}$.
\par
The Higgs flavor invariant (unrenormalized) form of the scalar Yukawa vertex $\Gamma^{o}(p,q,-p-q)$ is given by
\begin{equation}
\Gamma^{o}(p,q,-p-q)= \frac{<\phi(p) \sum \limits_{i=1}^{i=2} [\ h^{*i}(q)h^{i}(-p-q) + h^{*i}(-p-q) h^{i}(q) ]\ >}{D^{o}_{s}(p)D^{o}_{h}(q)D^{o}_{h}(-p-q)}
\label{eq:unrenormalizedYukawadef1}
\end{equation}
The $\nu^{th}$ component of momentum $p$ is given by
\begin{equation}
p_{\nu}(i) = \frac{2}{a} \sin \frac{\pi i}{N_{\nu}}
\label{eq:phymomentum1}
\end{equation} 
where $N_{\nu}$ and $i$ are lattice size in $\nu$ direction and the point on the lattice, respectively.
\par
One of the goals is to extract the representative functions of the Yukawa vertex using ML. Hence, the vertex was computed for every investigated set of parameters and in the entire two dimensional plane instead of limited field momentum settings \cite{Maas:2013aia}. However, to increase time efficiency while maintaining generality, the scalar field momentum is restricted to the time direction while allowing one of the two Higgs fields to have momentum values in the plane. The other Higgs field acquires the momentum dictated by the conservation principle as is apparent in equation \ref{eq:unrenormalizedYukawadef1}. The renormalized form of the vertex differs by a multiplicative constant which is calculated by the renormalization condition $\Gamma(\mu,\nu,-\mu-\nu)=\lambda$, with $\mu$ and $\nu$ being the field momentum values at the renormalization point. Hence, the renormalized vertex is given by the 
\begin{equation}
\Gamma(p,q,-p-q) = \frac{\lambda}{\Gamma^{o}(\mu,\nu,-\mu-\nu)} \Gamma^{o}(p,q,-p-q)
\label{eq:renormalizedYukawadef1}
\end{equation}
\subsubsection{Machine learning} \label{subsection:MachineLearning1}
The application of supervised machine learning \cite{MarslandML} on both the field propagators as well as the Yukawa vertex requires training of a suitable forms of the functions which are numerically accommodating, and physically meaningful if possible. For the field propagators, the following function is chosen for the scalar propagator:
\begin{equation}
 f_{ml.s}(p,m_{s},\alpha,\lambda,\gamma)= \frac{1}{p^{2}+m^{2}_{s}+\sum_{i,j,k,l,t =0}^{5} c_{ijklt} \alpha^{i}  (\frac{\lambda}{\Lambda})^{j} \gamma^{k} (\frac{p}{\Lambda})^{l} (\frac{m_{s}}{\Lambda})^{t} }
 \label{eq:mlfuncspr1}
\end{equation}
and the following function was used to train on the Higgs propagators
\begin{equation}
 f_{ml,h}(p,m_{s},\alpha,\lambda,\gamma)= \frac{1}{p^{2}+m^{2}_{h}+\sum_{i,j,k,l,t =0}^{5} \tilde{c}_{ijklt} \alpha^{i} (\frac{\lambda}{\Lambda})^{j} \gamma^{k} (\frac{p}{\Lambda})^{l} (\frac{m_{s}}{\Lambda})^{t} }.
 \label{eq:mlfunchpr1}
\end{equation}
The training function for Yukawa vertex requires further considerations as there are two independent field momenta. The following function was used for ML application.
\begin{equation}
F_{\Gamma}(m_{s},\alpha,\lambda,\beta,u,v)=  \lambda [ 1 + \sum_{i,j,k,l,m,n =0}^{i,j,k,l,m,n =4} c_{ijklmn} \tilde{m}_{s}^{i} \alpha^{j} \tilde{\lambda}^{k} \beta^{l} u^{m} v^{n} ]
\label{eq:mlfuncvtx1}
\end{equation}
where $u=a^{2}\underrightarrow{p}.\underrightarrow{q}$, $v=a^{2} \underrightarrow{q}.\underrightarrow{q}$, $\tilde{m}_{s}=a m_{s}$, and $\tilde{\lambda}_{s}=a \lambda_{s}$, where $a$ is the lattice spacing at the chosen point of the parameter space. $\underrightarrow{p}$ and $\underrightarrow{q}$ are the momentum vectors (in 4 dimensional Euclidean spacetime) of scalar and the Higgs fields, respectively. The vertex function is obtained under the renormalization condition $\Gamma(u(p,q),v(p,q))|_{p=\Lambda,q=\Lambda}=\lambda$ with $\Lambda$ being the highest momentum value along a dimension. It is important to mention here that the vertex was not projected on the expansion terms arising from perturbative expansion in the theory. Since the scalar Yukawa coupling is assumed to take any value, such expansions are not as meaningful as in perturbation theory. Hence, the variables $u$ and $v$ were deliberately chosen to expand the vertex in terms of scalar parameters formed by the field momenta respecting the conservation law.
\par
The cost function for each of the considered quantities is defined in terms of least square deviation between the results from LS and the prediction of the respective function being trained. As mentioned above, the functions are structured by the parameters of the theory which are not held fixed for the entire study, except in the tree level structure of the propagators, and field momenta.
\par
The implementation of the SML is as follows for the case of the scalar propagators: The training begins with all coefficients $c_{ijklt}$ set at zero resulting a certain value of the cost function value $f_{c}$ which may have a considerable magnitude because the starting point may not be the best description of the LS results. Each coefficient is systematically updated in search of a lower value of $f_{c}$. If one candidate value of $c_{ijklt}$ does not lower the cost function, update is not accepted and another candidate value is tested. As soon $f_{c}$ decreases, the candidate value is accepted for the $c_{ijklt}$ and the next coefficient is examined. In case no candidate for a certain $c_{ijklt}$ lowers the cost function, examination of another $c_{ijklt}$ begins. Once, all $c_{ijklt}$s are examined, it marks completion of one iteration. The process continues until either the number of iterations are exhausted (which is the case for our study), or the pre-set lowest value of the cost function is achieved.
\par
The process of numerically choosing a candidate value is as follows: A (non-vanishing and suitably large) value $\delta$ is chosen such that $c_{ijklt} \rightarrow c^{\prime}_{ijklt} = c_{ijklt} + \delta$. Cost function is computed against this change. If it results in a decrease of the function, the change is accepted and evaluation of another coefficient $c_{ijklt}$ starts. Else, $c^{\prime}_{ijklt} \rightarrow c_{ijklt}$ is performed and then a value $\frac{\delta}{10}$ is chosen such that $c_{ijklt} \rightarrow c^{\prime \prime}_{ijklt} = c_{ijklt} + \frac{\delta}{10}$ and $f_{c}$ is computed to see if it changes. If it lowers the cost function, the change is accepted, else another value $\frac{\delta}{100}$ is chosen. The process continues until (based upon the available resources) either all pre-decided choices for $\frac{\delta}{n}$ (where $n$ should be a large number) have been tried, or a decrease in the cost function occurs. The method, which seems unambiguously resource hungry and apparently time inefficient, was chosen appreciating the nonlinearities involved and to avoid pseudo-optimum point. The training follows what is termed as coordinate descent method \cite{nocedal2006numerical} in machine learning.
\par
For the case of the Higgs propagators, the same procedure is repeated using the function \ref{eq:mlfunchpr1}, while for the case of vertex it is the function \ref{eq:mlfuncvtx1} and there are two field momenta being considered instead of one.
\par
The lattice simulations, analysis of results from LS, and machine learning are performed in C++ environment, while ROOT CERN is used to generate the diagrams.
\section{Results} \label{section:Results}
\subsection{Field Propagators} \label{subsection:Propagators}
\begin{figure} [h]
 \centering
 \includegraphics[width = 1.0\textwidth]{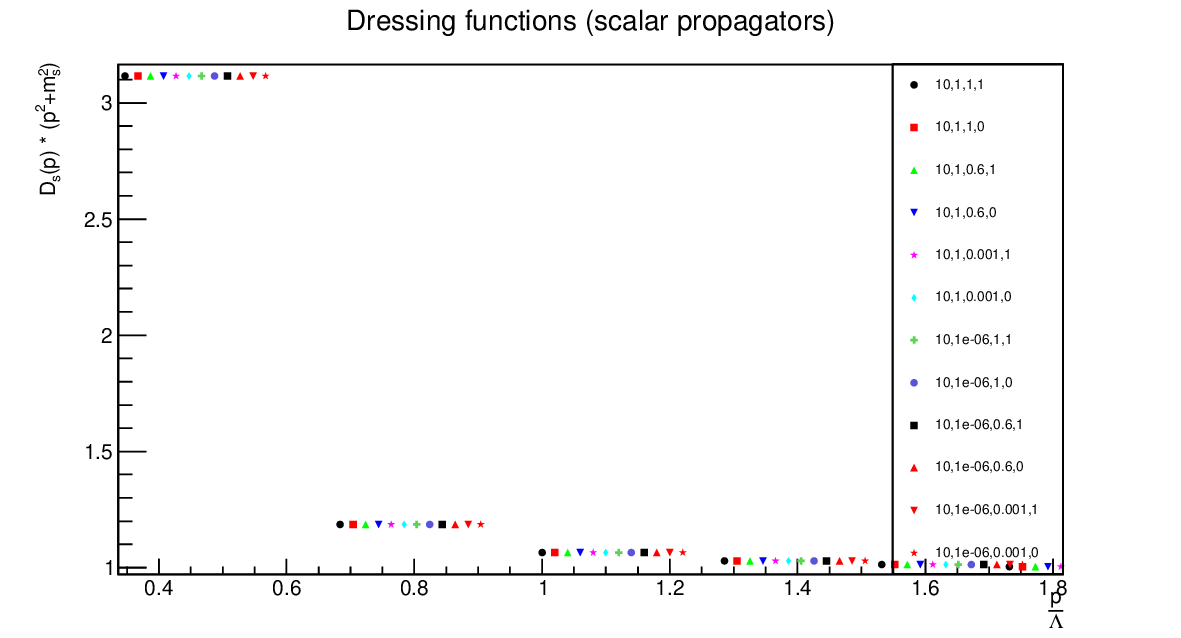}
 \caption{Dressing function of scalar propagator for scalar (lattice) mass $10$ against the parameters $(m_{s},\alpha,\lambda,\gamma)$ is shown.}
 \label{fig:s2p1}
\end{figure}
\begin{figure} [h]
 \centering
 \includegraphics[width = 1.0\textwidth]{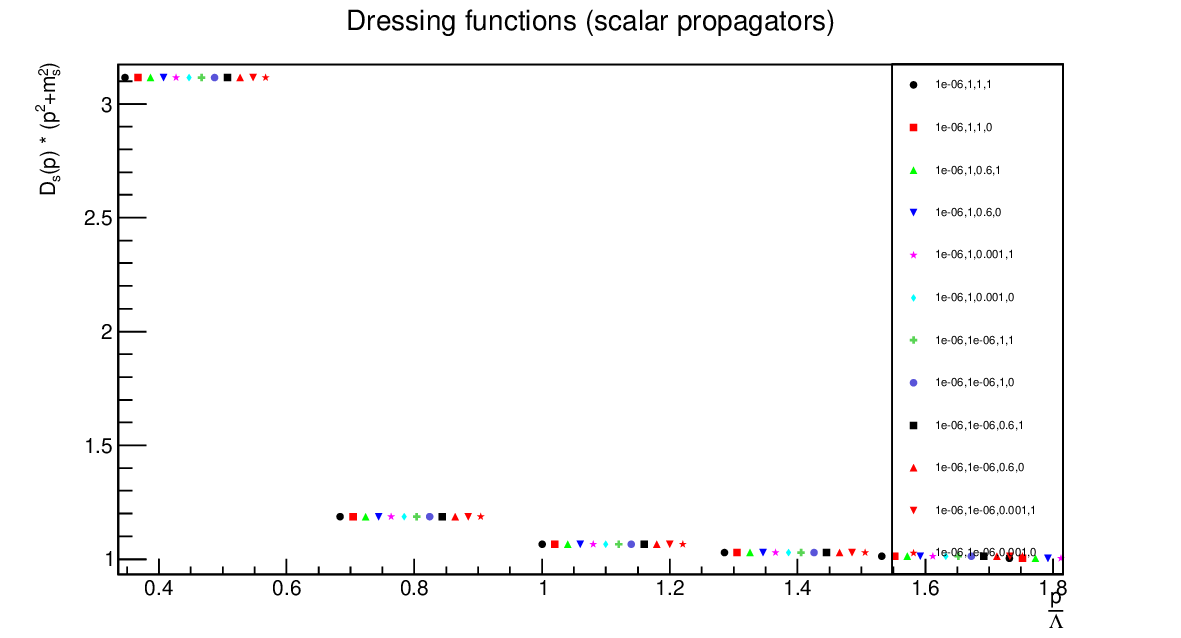}
 \caption{Dressing function of scalar propagator for scalar (lattice) mass $10^{-6}$ against the parameters $(m_{s},\alpha,\lambda,\gamma)$ is shown.}
 \label{fig:s2p2}
\end{figure}
\begin{figure} [h]
 \centering
 \includegraphics[width = 1.0\textwidth]{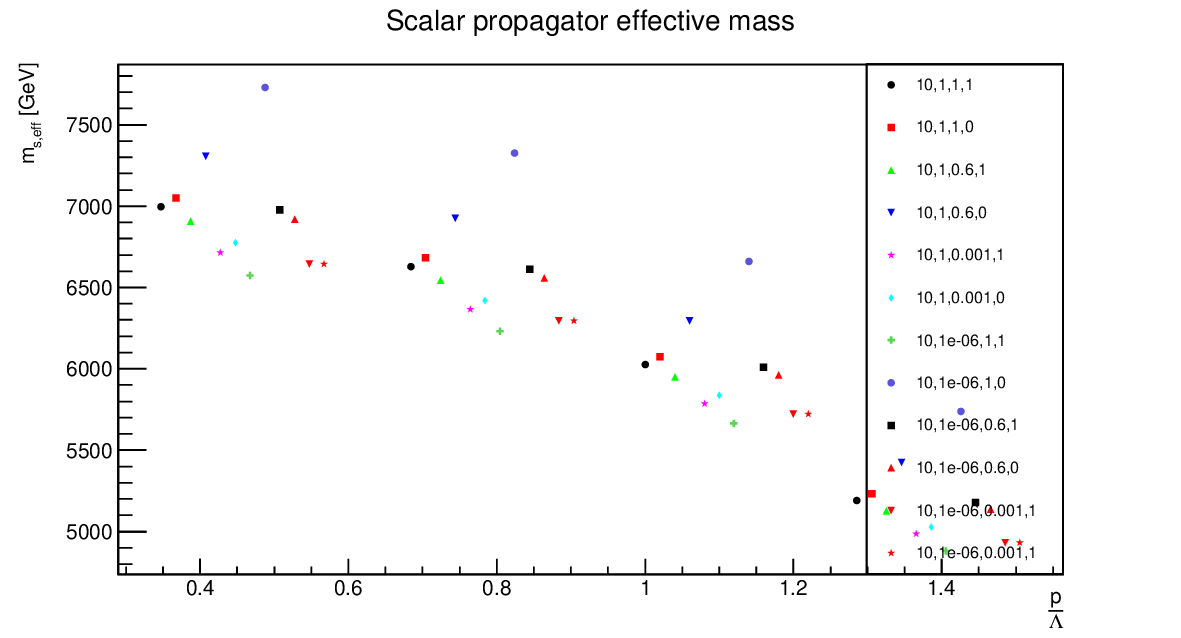}
 \caption{Scalar propagator effective mass of scalar (lattice) mass $10$ against the parameters $(m_{s},\alpha,\lambda,\gamma)$ is shown.}
 \label{fig:s2peffmass10p0}
\end{figure}
\begin{figure} [h]
 \centering
 \includegraphics[width = 1.0\textwidth]{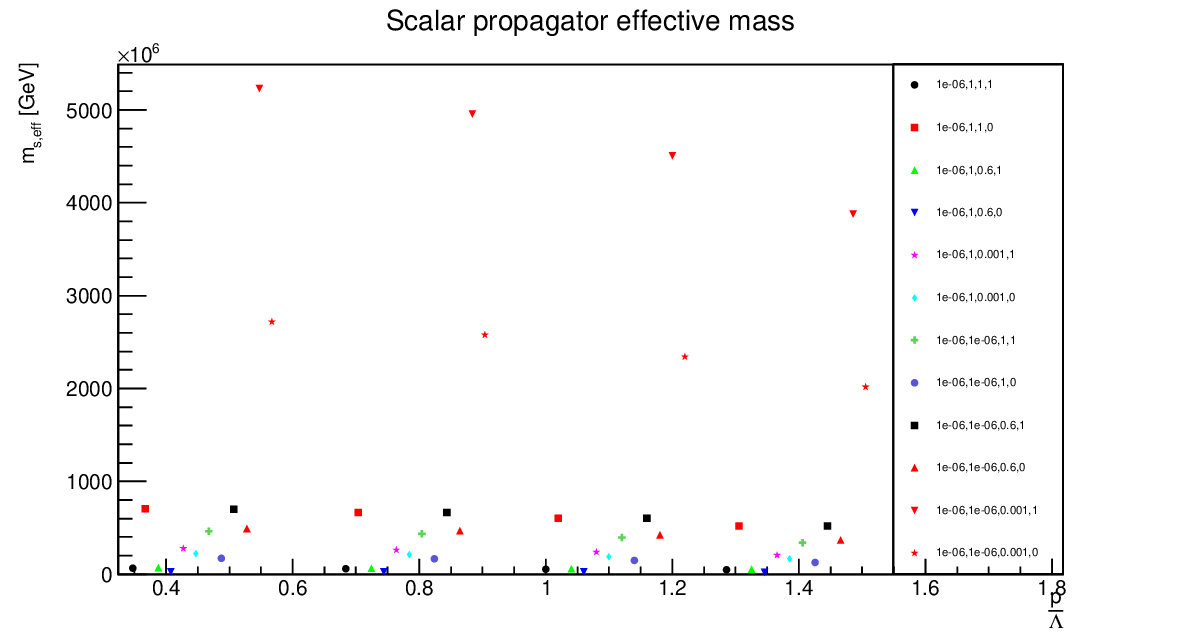}
 \caption{Scalar propagator effective mass of scalar (lattice) mass $10^{-6}$ against the parameters $(m_{s},\alpha,\lambda,\gamma)$ is shown.}
 \label{fig:s2peffmass0p000001}
\end{figure}
\begin{figure} [h]
 \centering
 \includegraphics[width = 1.0\textwidth]{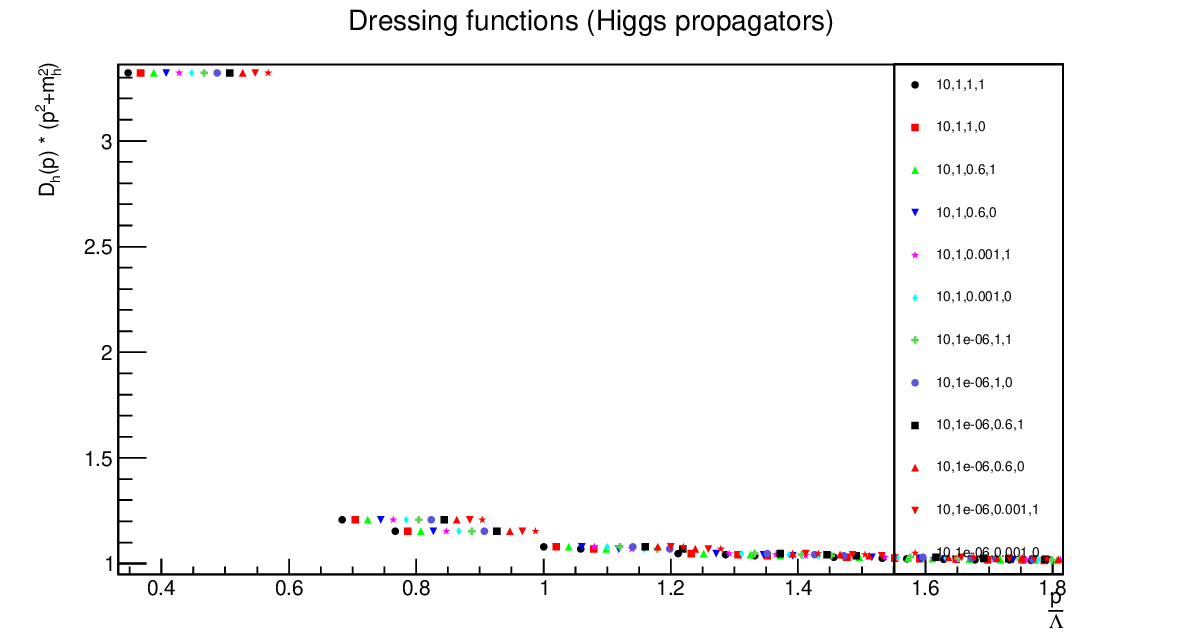}
 \caption{Dressing function of the Higgs propagator of scalar (lattice) mass $10$ against the parameters $(m_{s},\alpha,\lambda,\gamma)$ is shown.}
 \label{fig:h2p1}
\end{figure}
\begin{figure} [h]
 \centering
 \includegraphics[width = 1.0\textwidth]{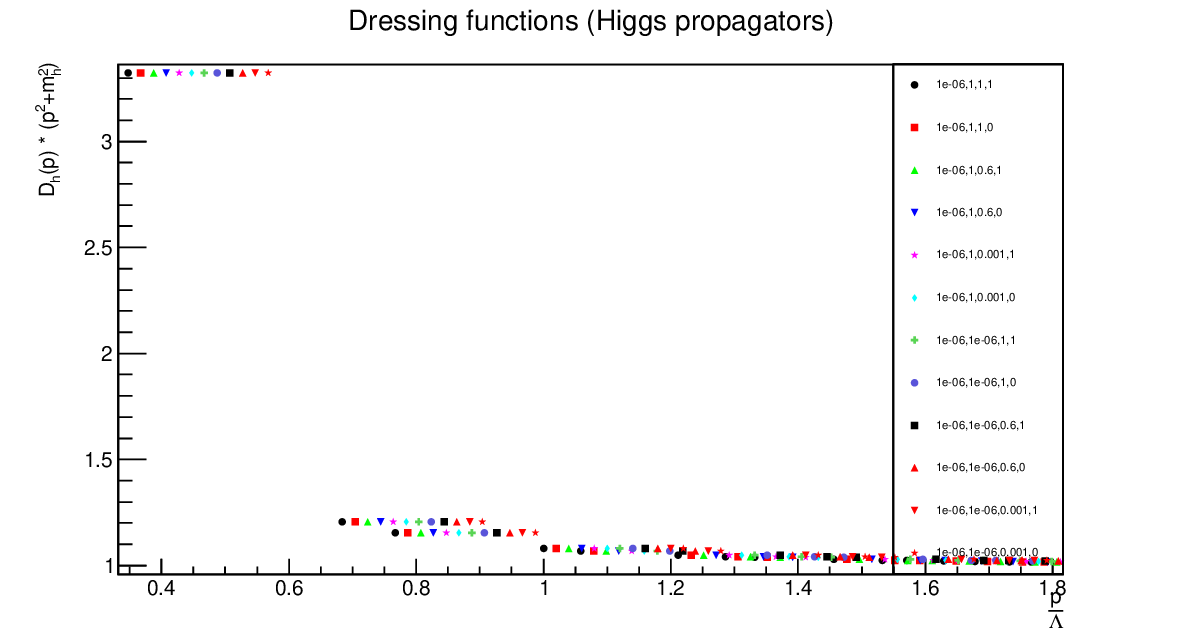}
 \caption{Dressing function of the Higgs propagator for scalar (lattice) mass $10^{-6}$ against the parameters $(m_{s},\alpha,\lambda,\gamma)$ is shown.}
 \label{fig:h2p2}
 \end{figure}
\begin{figure} [h]
 \centering
 \includegraphics[width = 1.0\textwidth]{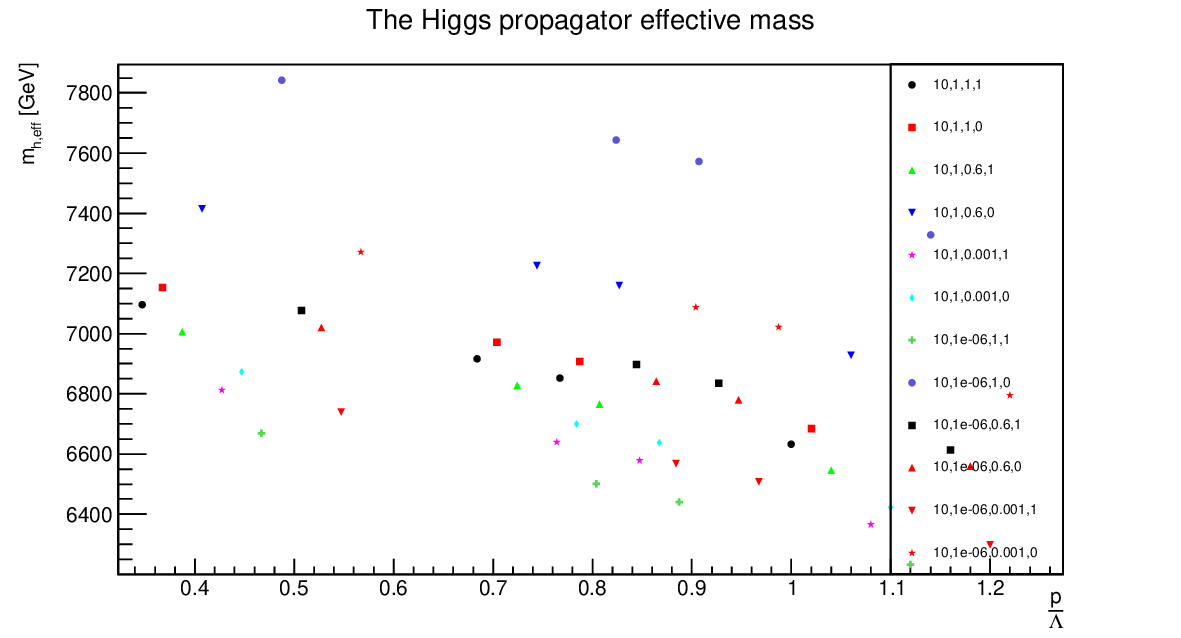}
 \caption{Higgs propagator effective mass for scalar (lattice) mass $10$ against the parameters $(m_{s},\alpha,\lambda,\gamma)$ is shown.}
 \label{fig:h2peffmass10p0}
\end{figure}
\begin{figure} [h]
 \centering
 \includegraphics[width = 1.0\textwidth]{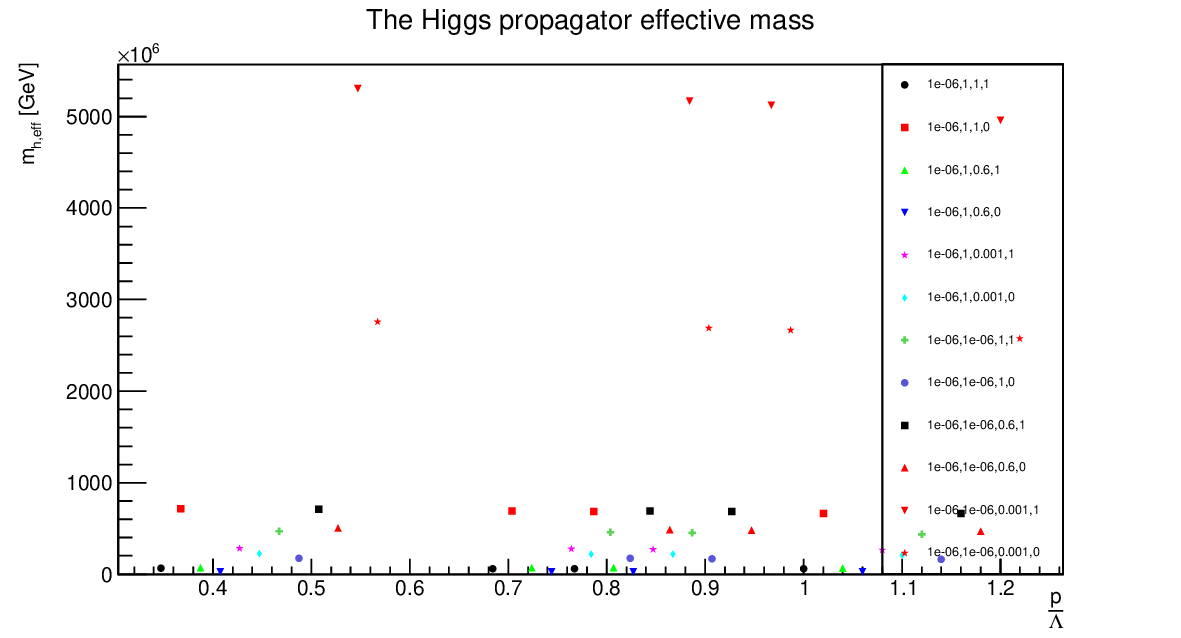}
 \caption{Higgs propagator effective mass for scalar (lattice) mass $10^{-6}$ against the parameters $(m_{s},\alpha,\lambda,\gamma)$ is shown.}
 \label{fig:h2peffmass0p000001}
\end{figure}
\begin{figure} [h]
 \centering
 \includegraphics[width = 1.0\textwidth]{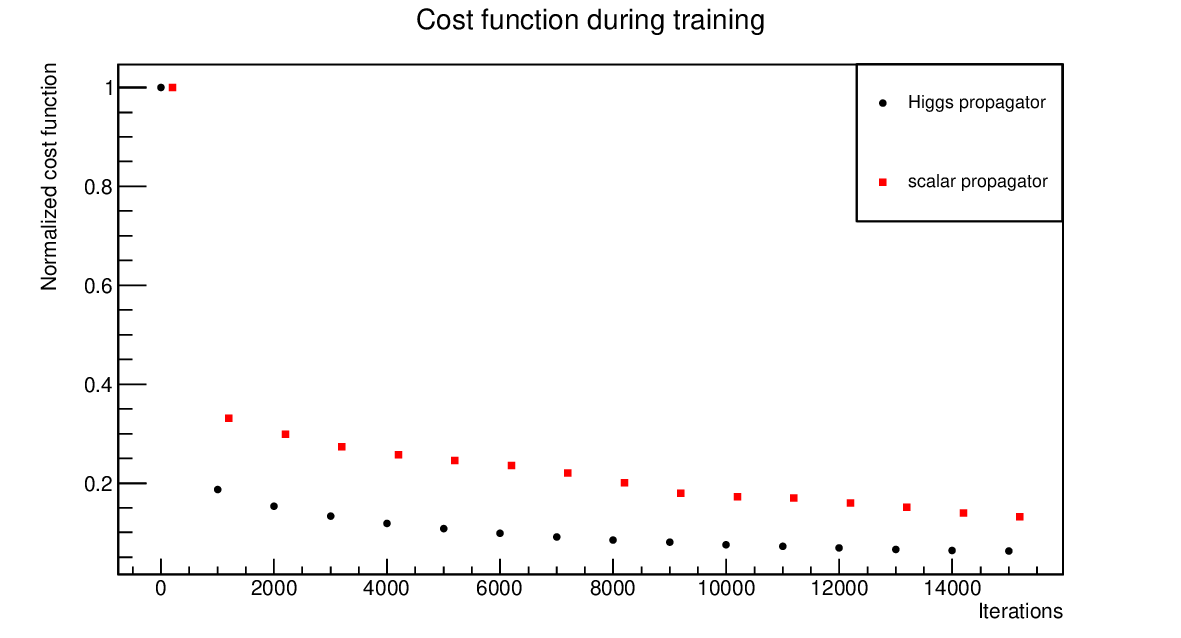}
 \caption{Cost functions of the field propagators are shown against iterations performed during machine learning.}
 \label{fig:propscostfunctions}
\end{figure}
\begin{figure} [h]
 \centering
 \includegraphics[width = 1.0\textwidth]{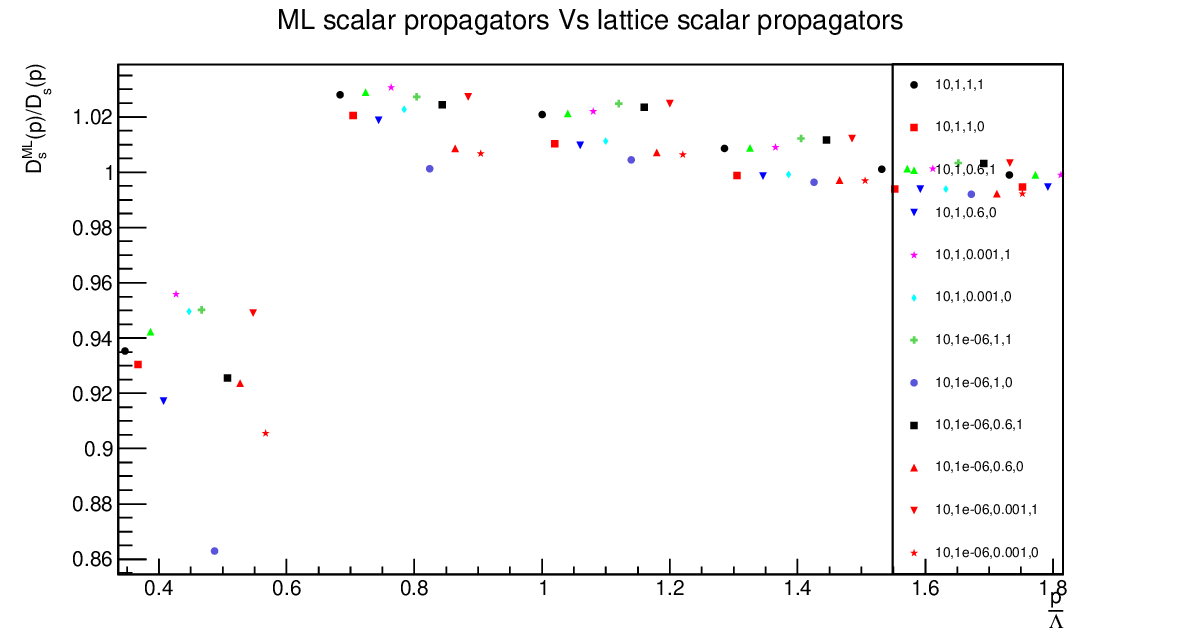}
 \caption{Ratio between ML generated and lattice computed Scalar propagators for scalar (lattice) mass $10$ against the parameters $(m_{s},\alpha,\lambda,\gamma)$ is shown.}
 \label{fig:s2prat1}
\end{figure}
\begin{figure} [h]
 \centering
 \includegraphics[width = 1.0\textwidth]{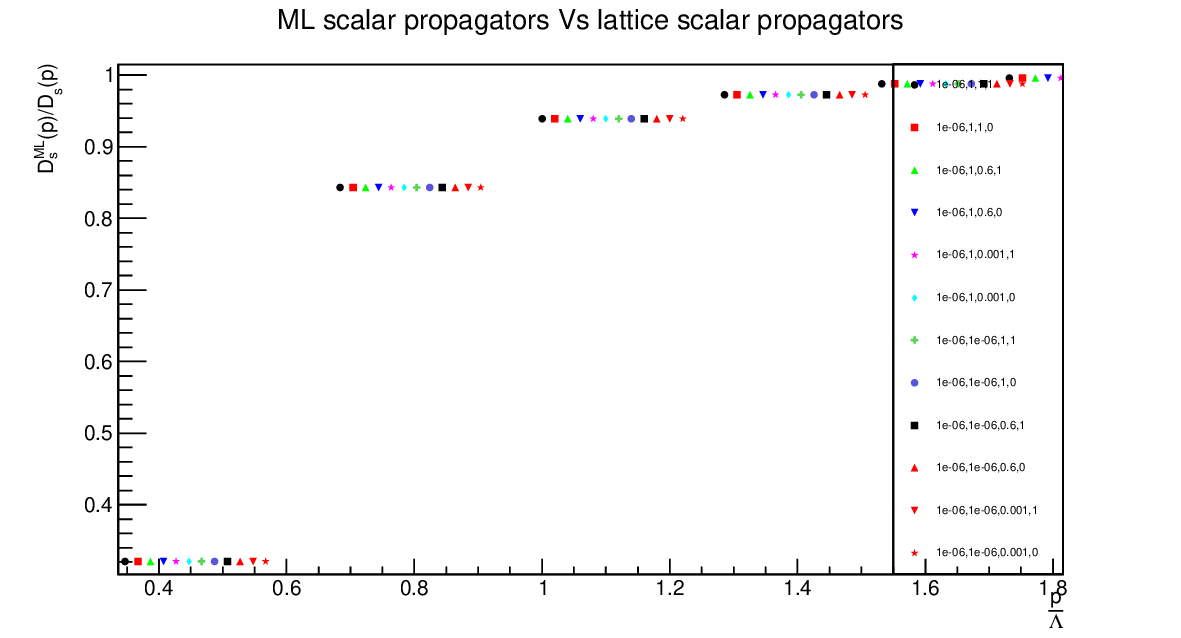}
 \caption{Ratio between ML generated and lattice computed scalar propagators for scalar (lattice) mass $10^{-6}$ against the parameters $(m_{s},\alpha,\lambda,\gamma)$ is shown.}
 \label{fig:s2prat2}
\end{figure}
\begin{figure} [h]
 \centering
 \includegraphics[width = 1.0\textwidth]{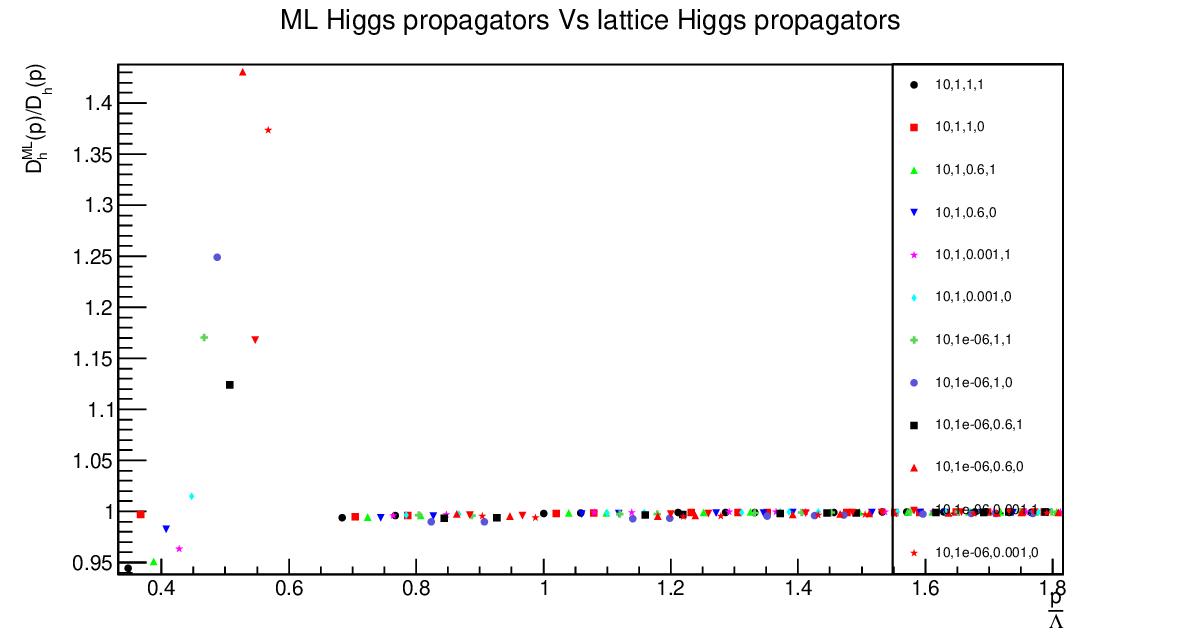}
 \caption{Ratio between ML generated and lattice computed Higgs propagators for scalar (lattice) mass $10$ against the parameters $(m_{s},\alpha,\lambda,\gamma)$ is shown.}
 \label{fig:h2prat1}
\end{figure}
\begin{figure} [h]
 \centering
 \includegraphics[width = 1.0\textwidth]{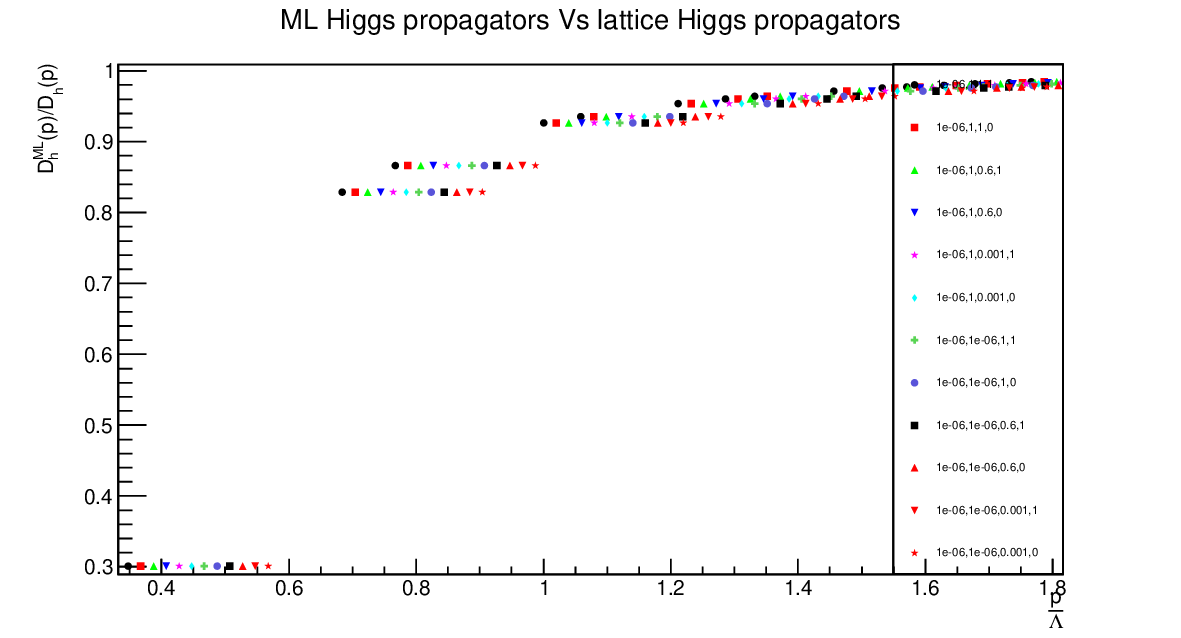}
 \caption{Ratio between ML generated and lattice computed Higgs propagators for scalar (lattice) mass $10^{-6}$ against the parameters $(m_{s},\alpha,\lambda,\gamma)$ is shown.}
 \label{fig:h2prat2}
\end{figure}
\begin{figure} [h]
 \centering
 \includegraphics[width = 1.0\textwidth]{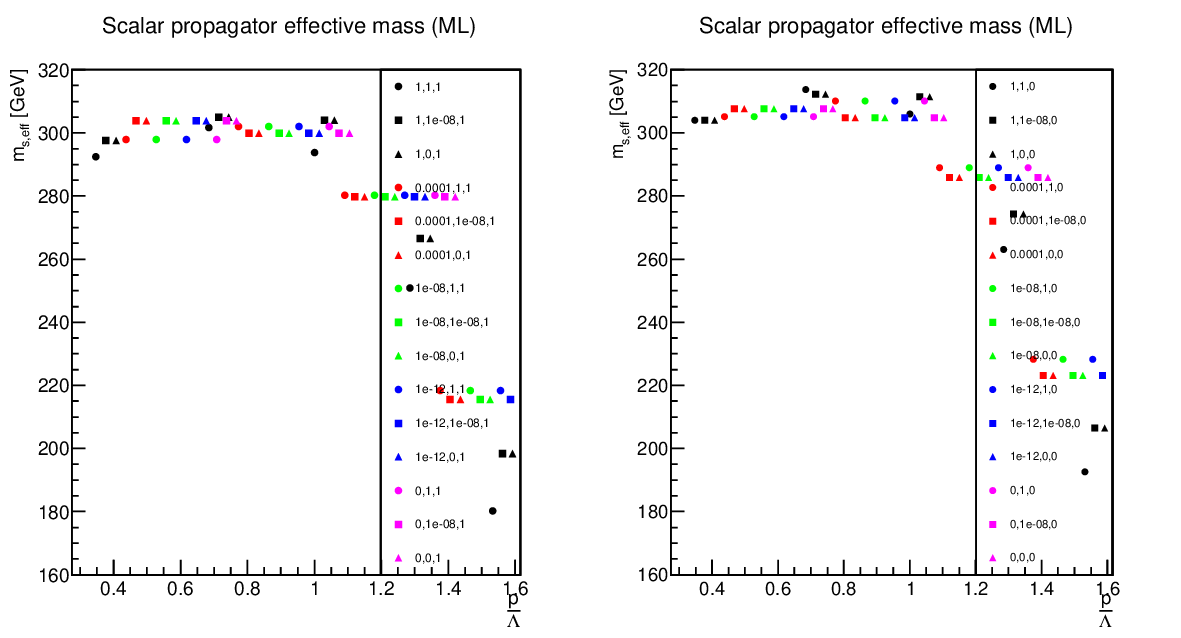}
 \caption{Scalar effective mass at $m_{s}=0$ and $\Lambda=1TeV$ is shown.}
 \label{fig:seffmass1TeV}
\end{figure}
\begin{figure} [h]
 \centering
 \includegraphics[width = 1.0\textwidth]{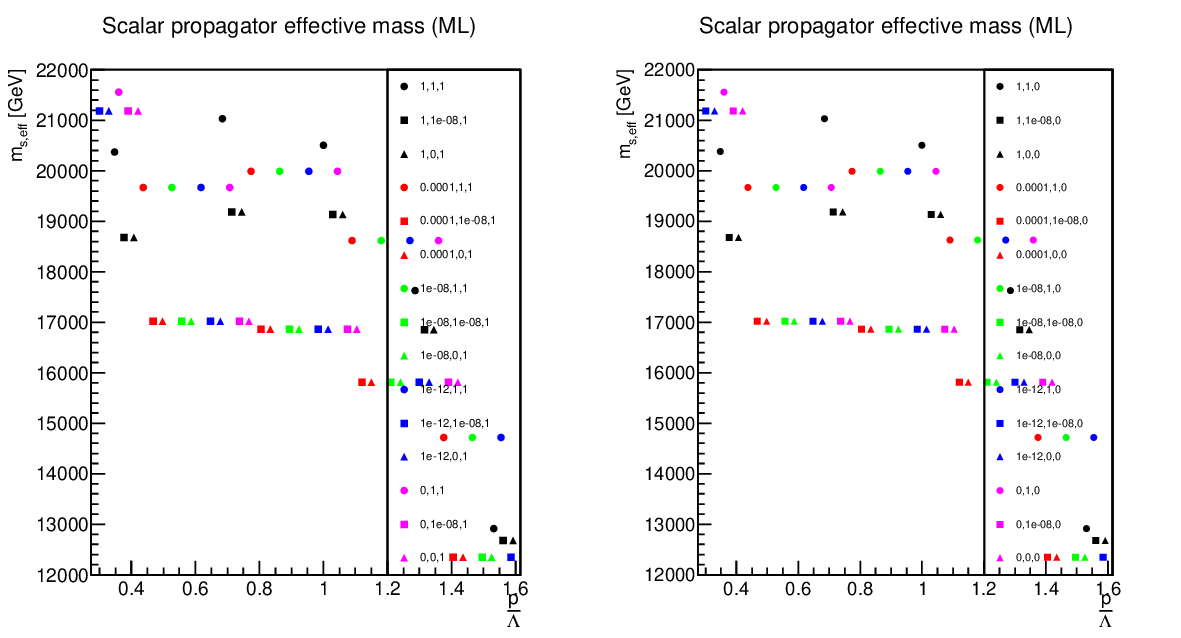}
 \caption{Scalar effective mass at $m_{s}=0$ and $\Lambda=10^{3}TeV$ is shown.}
 \label{fig:seffmass1000TeV}
\end{figure}
\begin{figure} [h]
 \centering
 \includegraphics[width = 1.0\textwidth]{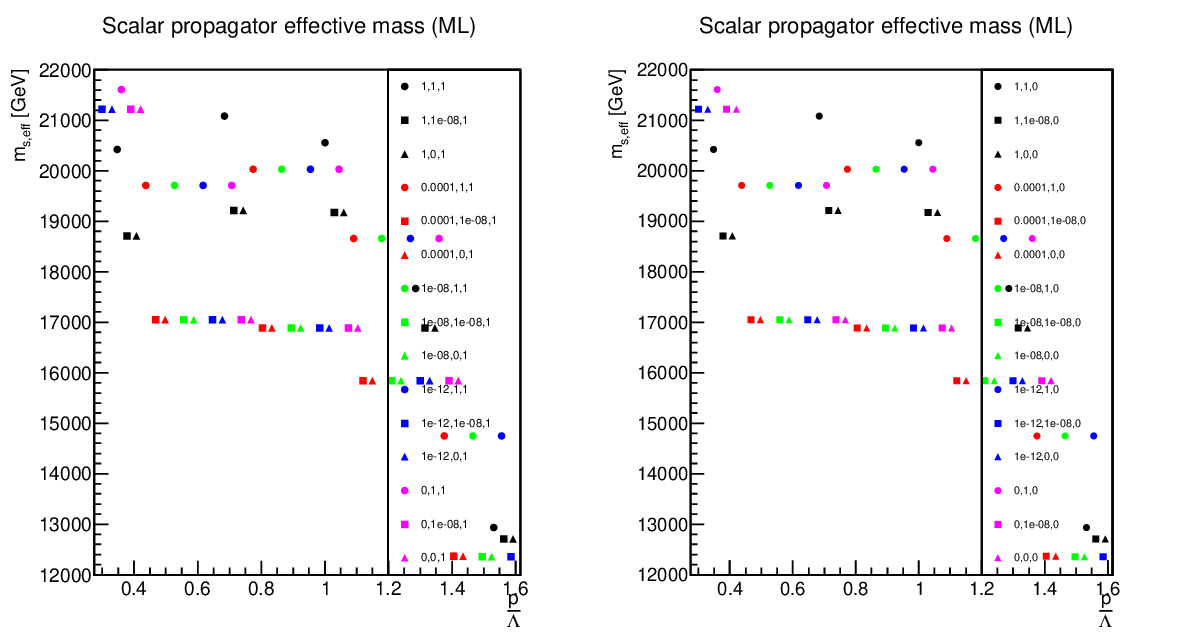}
 \caption{Scalar effective mass at $m_{s}=0$ and $\Lambda=10^{6}TeV$ is shown.}
 \label{fig:seffmass1000000TeV}
\end{figure}
\begin{figure} [h]
 \centering
 \includegraphics[width = 1.0\textwidth]{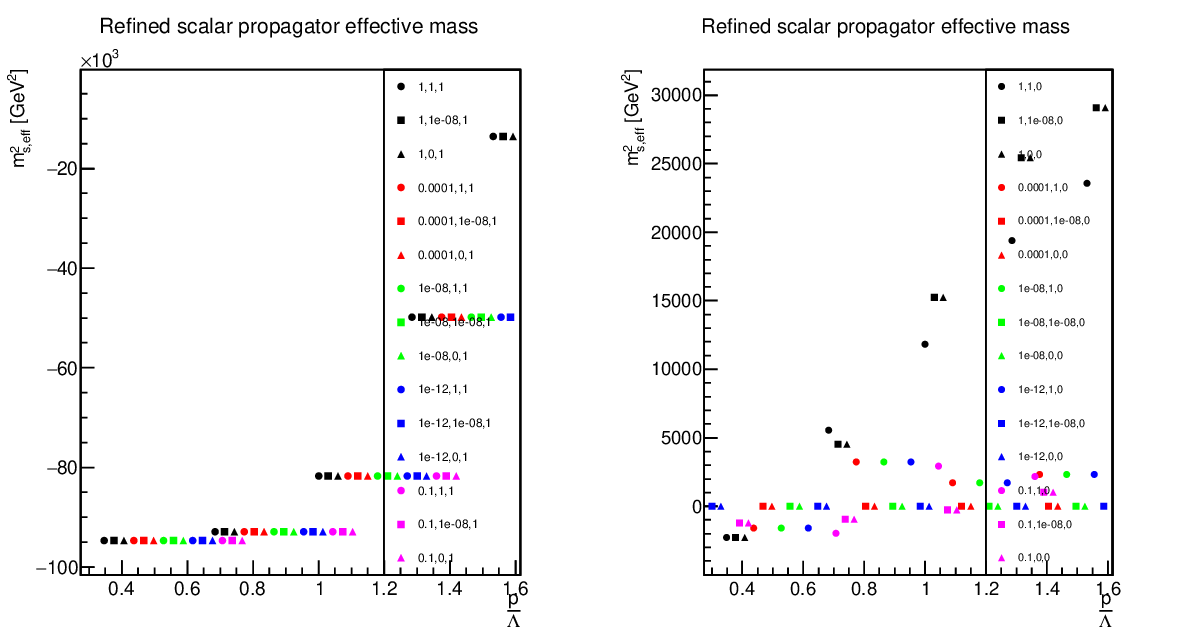}
 \caption{Scalar effective mass at $m_{s}=0$ and $\Lambda=1TeV$ is shown.}
 \label{fig:relseffmass1TeV}
\end{figure}
\begin{figure} [h]
 \centering
 \includegraphics[width = 1.0\textwidth]{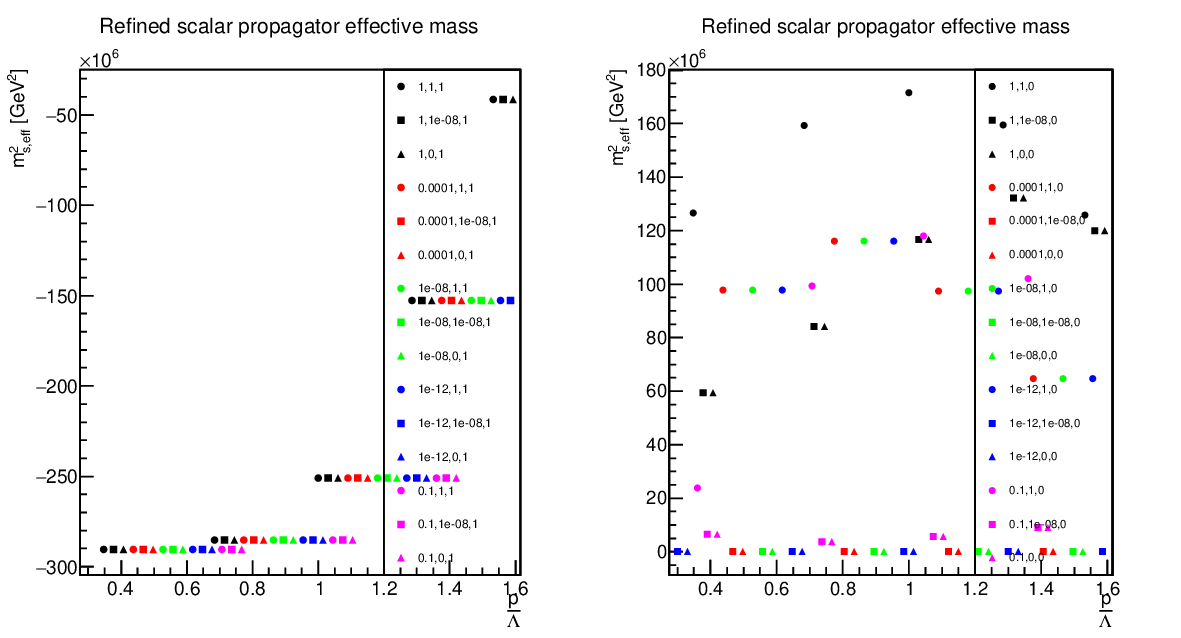}
 \caption{Scalar effective mass at $m_{s}=0$ and $\Lambda=10^{6}TeV$ is shown.}
 \label{fig:relseffmass1000000TeV}
\end{figure}
The model is found to have a wide variety of scales (regulators) for not a very large size of the lattice \footnote{The problem can be partially alleviated on very large lattice size, which is not possible due to computation of the Yukawa vertex on the entire plane.}, see Appendix E. It renders limited access to the infrared region of the model. The application of machine learning to the propagators provides a possibility of cautious extrapolation in the infrared region while portraying a large picture of these correlation functions in the parameter space, while an obvious by product is availability of the results for studies involving Dyson Schwinger equations.
\par
The dressing functions of the scalar propagators are shown in Figures \ref{fig:s2p1}-\ref{fig:s2p2}. Presence of small mutual deviations is an indicative of similar qualitative behavior. However, a remarkable feature is rise of the propagators in infrared region compared to the tree level structure in Euclidean spacetime which is an unambiguous demonstration of nonperturbative effects that may not be captured by perturbation theory. It supports speculating formation of a pole originating in the self energy term. To further probe the behavior, effective mass $m_{s,eff}$ for the scalar propagator is computed, shown in Figures \ref{fig:s2peffmass10p0}-\ref{fig:s2peffmass0p000001}. The quantity $m_{s,eff}$ is defined by 
\begin{equation}
 m_{s,eff} = p^{2} - \frac{1}{D_{s}(p)}
 \label{eq:defSeffmass}
\end{equation} 
while the corresponding magnitude of the error is calculated by
\begin{equation}
E[m_{s,eff}] = \frac{E[D_{s}(p)]}{D^{2}_{s}(p)}.
\label{eq:defSeffmassE}
\end{equation}
The results strongly point towards presence of pole in the scalar propagator throughout the parameter space. A peculiar sensitivity on $m_{s}$ is found inversely related to the possible pole, i.e. $m_{s}=10$ favors the pole in the vicinity of $7.5$ TeV while decreasing the scalar (lattice) mass raises the pole to a considerably higher poles due to mostly large regulator. Such a large modification of the tree level structure, particularly at stronger coupling, can not be accessible through perturbation theory. Hence, we witness nonperturbative effects in the scalar propagators in the guise of dynamical generation of pole-like behavior.
\par
A remarkable feature is that Higgs propagator behaves similarly despite bearing a different symmetry, see Figures \ref{fig:h2p1}-\ref{fig:h2p2}. The observations underscores the sensitivity of the scalar (bare) mass in the model which eventually effects the scale (regulator) in the model, and the effective mass shown in Figures \ref{fig:h2peffmass10p0} and \ref{fig:h2peffmass0p000001} and defined below:
\begin{equation}
 m_{h,eff} = p^{2} - \frac{1}{D_{h}(p)}
 \label{eq:defHeffmass}
\end{equation}
while the corresponding magnitude of the error is calculated by
\begin{equation}
E[m_{h,eff}] = \frac{E[D_{h}(p)]}{D^{2}_{h}(p)}.
\label{eq:defHeffmassE}
\end{equation}
To complement the lattice results, machine learning is used on the lattice results against all considered points, see Table \ref{table:masses-n-couplings}, in the parameter space in order to analytically capture the core of the propagators. For this reason, nonlinear regression method is chosen instead of other interesting approaches including artificial neural network (ANN) \cite{MarslandML}. An advantage is control over the mathematical structure of the function to be trained which facilitates in understanding the underlying physics. These functions are given in equations \ref{eq:mlfuncspr1} and \ref{eq:mlfunchpr1} for the scalar and the Higgs propagators, respectively. Evidently, it is the self-energy term in the propagators which is trained. The cost functions for the respective trainings are shown in Figure \ref{fig:propscostfunctions} while the definition of these functions for scalar and the Higgs propagators are respectively given below:
\begin{equation}
f_{cost,s} = \sum_{ijkln} |\ f_{ml,s}(p_{i},m_{s,j},\alpha_{k},\lambda_{l},\gamma_{n}) - D_{s}(p_{i},m_{s,j},\alpha_{k},\lambda_{l},\gamma_{n}) |\
\label{eq:costfuncspr1}
\end{equation} 
\begin{equation}
f_{cost,h} = \sum_{ijkln} |\ f_{ml,h}(p_{i},m_{s,j},\alpha_{k},\lambda_{l},\gamma_{n}) - D_{h}(p_{i},m_{s,j},\alpha_{k},\lambda_{l},\gamma_{n}) |\
\label{eq:costfunchpr1}
\end{equation} 
where $p_{i}$ are the physical momentum values, see equation \ref{eq:phymomentum1}, while $m_{s,j}$, $\alpha_{k}$, $\lambda_{l}$, and $\gamma_{n}$ are the values of the corresponding parameters in lattice units \footnote{Since the Higgs (bare) mass was kept constant, it was not included explicitly in the definitions of the cost functions.}. Due to faster convergence for the case of the Higgs propagator, 3 times longer training for scalar propagator is performed in order to achieve similar precision for both cases. The comparison between the LS and ML results are represented in Figures \ref{fig:s2prat1} and \ref{fig:s2prat2} for scalar propagators, and in Figures \ref{fig:h2prat1} and \ref{fig:h2prat2} for the Higgs propagators. The figures depict the accuracy achieved at the end of machine learning. The deviations (from 1) may arise due to limited capacity of the functions used for training, the lattice artifacts, and statistical noise. A reasonable compatibility of ML results with the LS results is observed over most of the points, except the lowest momentum values, against most of the points in the model's parameter space.
\par
To analyze ML results, first we take a brutal assumption that the trained function in equation \ref{eq:mlfuncspr1} has sufficiently captured the underlying physics and the deviations between the ML and LS results, presented in Figures \ref{fig:h2prat1}-\ref{fig:h2prat2} and \ref{fig:s2prat1}-\ref{fig:s2prat2}, are solely due to lattice artifacts. As the assumption legitimizes drastic extrapolations, we consider the case $m_{s}=0$ against different values of regulator $\Lambda$ in order to examine the dynamically generated values of the pole in the scalar field propagator which by definition is a non-perturbative phenomenon, though formally known in QCD interactions \cite{Finger:1981gm}.
\par
The effective scalar masses are shown in Figures \ref{fig:seffmass1TeV}-\ref{fig:seffmass1000000TeV}. The first observation is lifting of the effective mass from the vicinity of $300$ GeV, at $1$ TeV regulator, to the vicinity of $20$ TeV at $10^{6}$ TeV regulator, as the value of regulator increases in the parameter space. It provides an estimate of the dynamically developed pole, in the scalar propagator. In addition, Figures \ref{fig:seffmass1TeV}-\ref{fig:seffmass1000000TeV} also indicate a subtle role of the Yukawa coupling, i.e. higher the coupling strength results decrease in the magnitude of the effective mass, despite that the effect is weak and disappears as the effective mass rises to comparatively higher values.
\par
However, the ML results requires further probing due to observation of effective mass even when the two fields decouple at all couplings involving scalar singlet set to 0. We proceed by refining the effective scalar mass by redefining it in the following.
\begin{equation}
 m^{2}_{s,eff} = p^{2} - \frac{1}{D_{s}(p)} - m^{2}_{s,eff,0}
 \label{eq:difSeffmassref}
\end{equation} 
where the quantity $m^{2}_{s,eff,0}$ is defined as
\begin{equation}
m^{2}_{s,eff,0} = p^{2} - \frac{1}{D_{s,0}(p)}.
\label{eq:difSeffmass0ref}
\end{equation} 
with $D_{s,0}(p)$ is the scalar renormalized ML propagator with all couplings involving scalar singlet set at 0. The Figures \ref{fig:relseffmass1TeV}-\ref{fig:relseffmass1000000TeV} describe the behavior of refined form of $m^{2}_{s,eff}$ in which \textit{refined} effective mass from scalar propagator is plotted. We observe significant deviations from the earlier defined form for the effective mass against a number of points in the parameter space.
\par
The role of the Yukawa coupling is prominent at both regulators which establishes undeniable role of the corresponding interaction vertex. The effects of other interactions is insignificant at $1$ TeV cutoff, while the Yukawa coupling enforces a peculiar trend which introduces stability in the IR region, see Figures \ref{fig:relseffmass1TeV}-\ref{fig:relseffmass1000000TeV}. As the Yukawa coupling decreases, sensitivity on other couplings emerges with a distinct effects of the highest coupling considered, i.e. the unit magnitude. The role of the regulator (lattice spacing) is also significant which emphasizes sensitivity to the parameter space up to the artifacts steming from various origins.
\par
The ML results and their analysis unambiguously support existence of pole in machine learnt scalar propagator, as is evident in Figures \ref{fig:seffmass1TeV}-\ref{fig:seffmass1000000TeV} and \ref{fig:relseffmass1TeV}-\ref{fig:relseffmass1000000TeV}. Thus, based upon a reasonable agreement between the ML and LS propagators, shown in Figures \ref{fig:s2prat1} and \ref{fig:s2prat2}, concluding the same for the LS scalar propagator is reasonable. Furthermore, the role of the Yukawa coupling strongly suggests that, despite a different symmetry, the Yukawa vertex may not be a null function.
\par
The phenomenon of dynamical mass generation is non-perturbative in nature and is mostly relevant to fermionic interactions \cite{Curtis:1992gm,Chang:2024xjd,Berbig:2024uwm,ptz061} with rare exceptions \cite{Kan:2023lah}. Existence of dynamics in the model bearing semblance to the phenomenon is remarkable. A true extrapolation of the ML based results deep into IR region, however, may not be as reliable. However, a safer strategy is to search for clear signals in the propagators which point towards existence of poles, which is the adopted approach for the study.
\par
\subsection{Scalar Yukawa Vertex} \label{section:Vertex}
\begin{figure} [h]
 \centering
 \includegraphics[width = 1.0\textwidth]{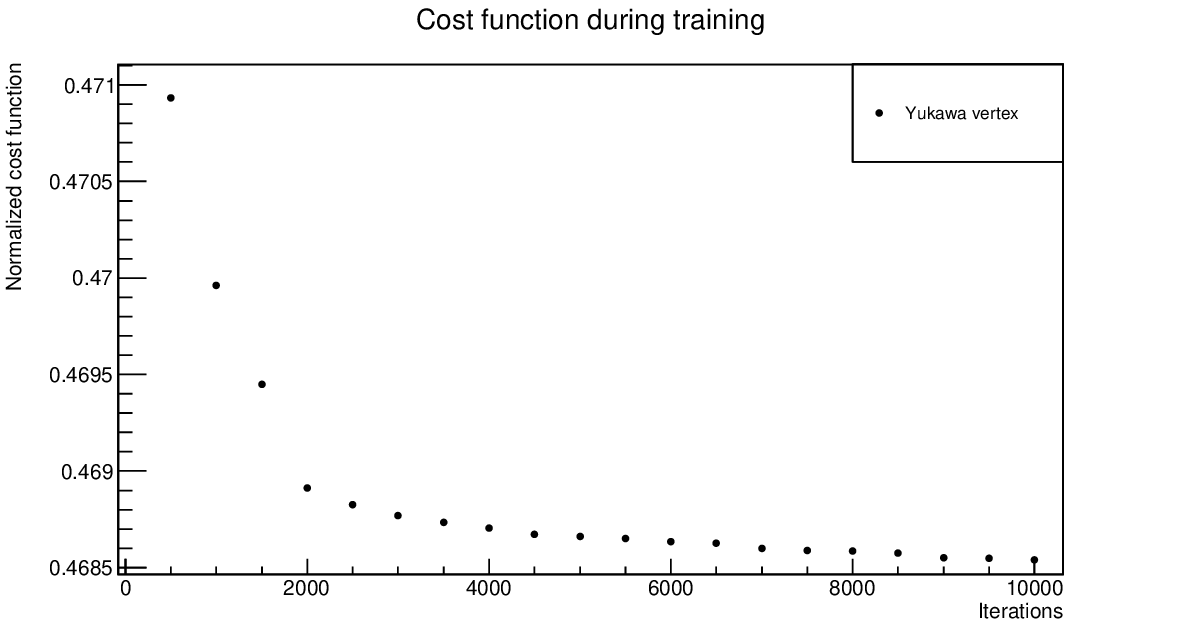}
 \caption{Cost function of machine learning the Yukawa vertex is shown.}
 \label{fig:vtxcostfunc1}
\end{figure}
\begin{figure} [h]
 \centering
 \includegraphics[width = 1.0\textwidth]{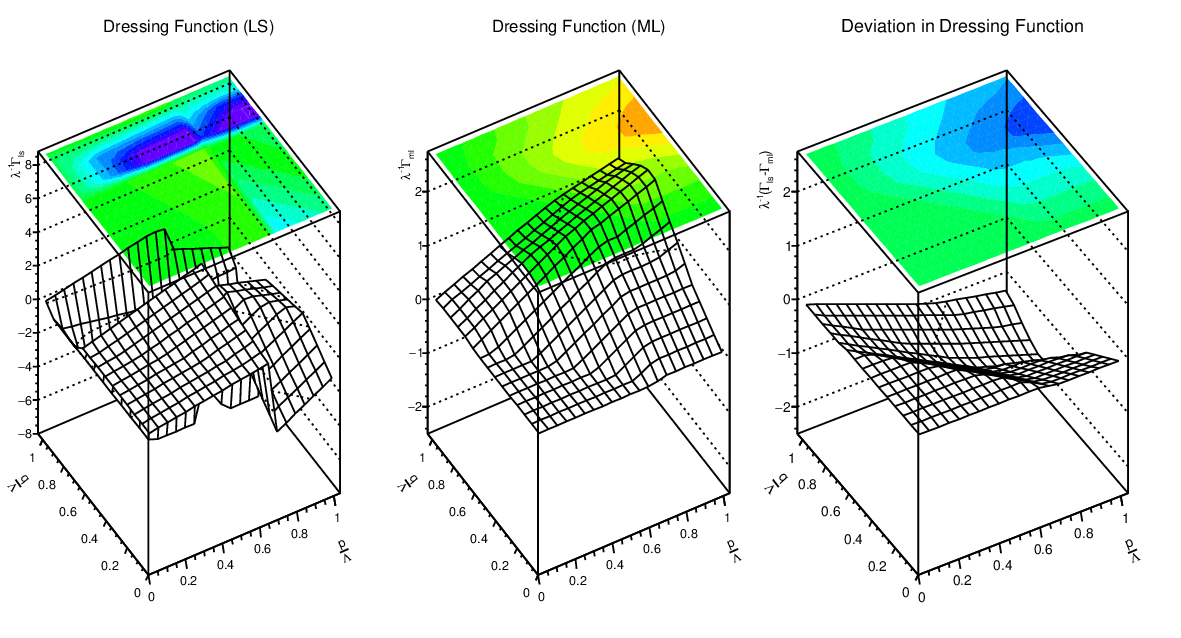}
 \caption{Yukawa vertex dressing function for $m_{s}=10.0$, $\alpha=1.0$, $\lambda=1.0$, and $\gamma=0.1$ is plotted. $\Gamma_{ls}$, $\Gamma_{ml}$, and $\Gamma_{ls}-\Gamma_{ml}$ represent the vertex from lattice simulations, the vertex from the results of SML, and the difference between the two results.}
 \label{fig:vertex1}
\end{figure}
\begin{figure} [h] 
 \centering
 \includegraphics[width = 1.0\textwidth]{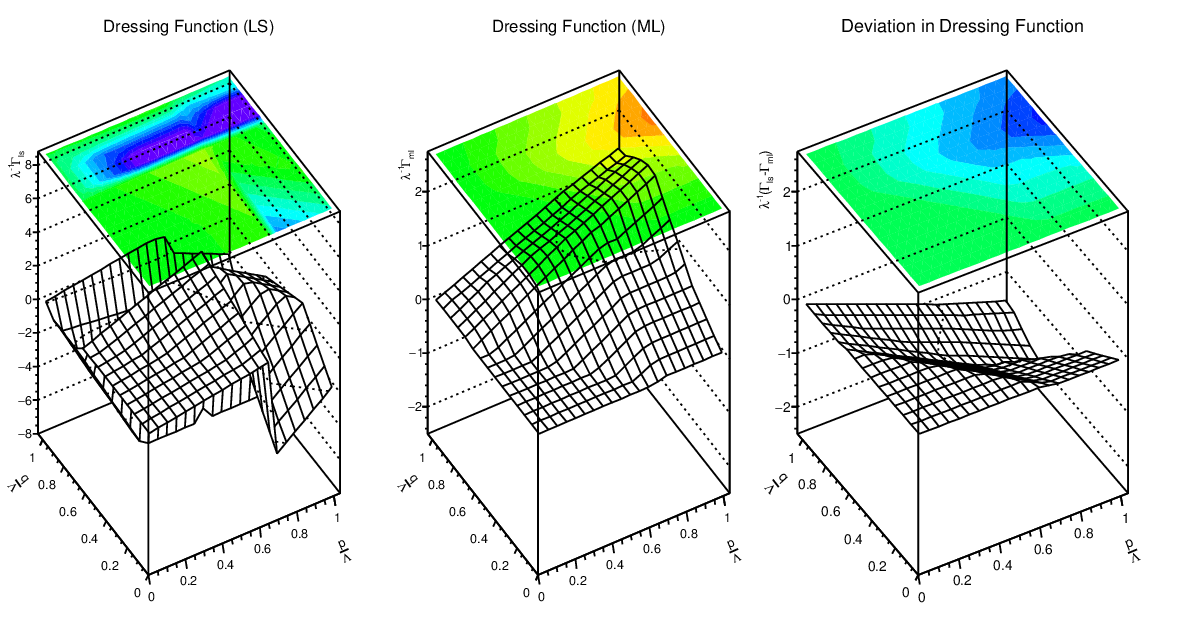}
 \caption{Yukawa vertex dressing function for $m_{s}=10.0$, $\alpha=1.0$, $\lambda=0.001$, and $\gamma=0.0$ is plotted. $\Gamma_{ls}$, $\Gamma_{ml}$, and $\Gamma_{ls}-\Gamma_{ml}$ represent the vertex from lattice simulations, the vertex from the results of SML, and the difference between the two results.}
 \label{fig:vertex2}
\end{figure}
\begin{figure} [h] 
 \centering
 \includegraphics[width = 1.0\textwidth]{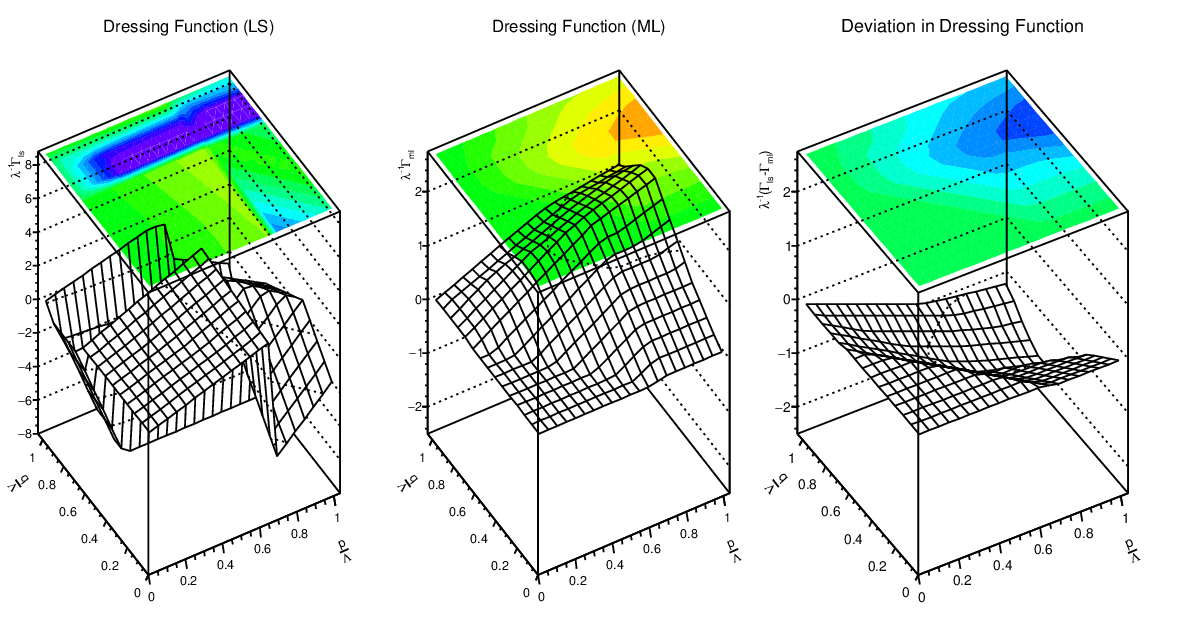}
 \caption{Yukawa vertex dressing function for $m_{s}=10^{-6}$, $\alpha=1.0$, $\lambda=1.0$, and $\gamma=1.0$ is plotted. $\Gamma_{ls}$, $\Gamma_{ml}$, and $\Gamma_{ls}-\Gamma_{ml}$ represent the vertex from lattice simulations, the vertex from the results of SML, and the difference between the two results.}
 \label{fig:vertex3}
\end{figure}
\begin{figure} [h] 
 \centering
 \includegraphics[width = 1.0\textwidth]{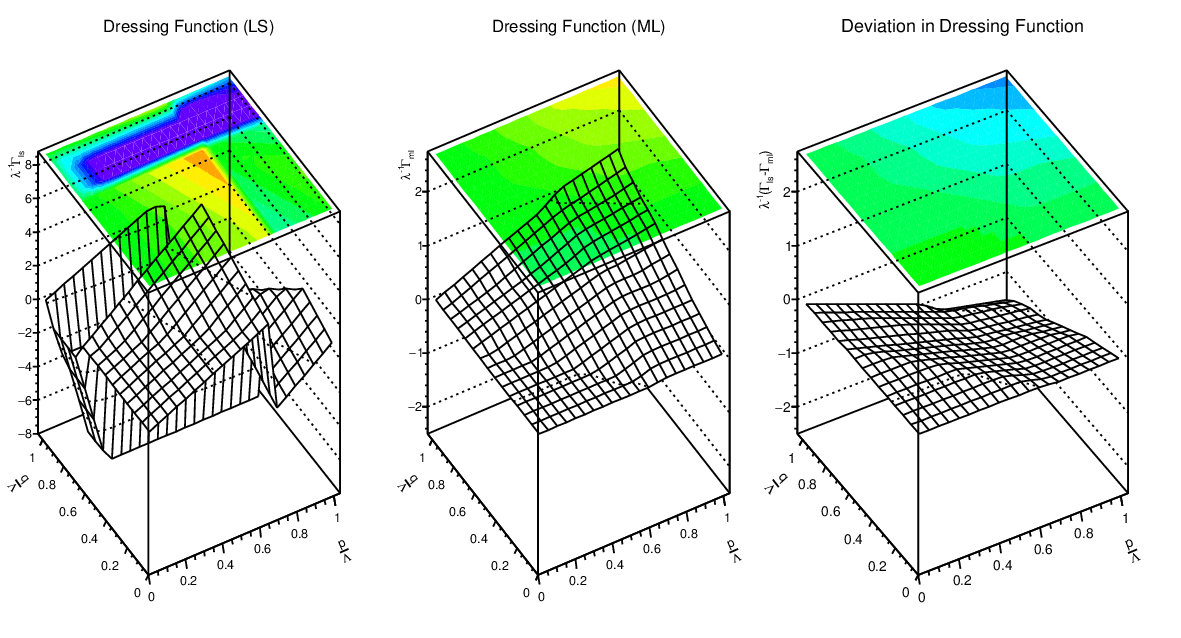}
 \caption{Yukawa vertex dressing function for $m_{s}=10^{-6}$, $\alpha=10^{-12}$, $\lambda=10^{-3}$, and $\gamma=0.0$ is plotted. $\Gamma_{ls}$, $\Gamma_{ml}$, and $\Gamma_{ls}-\Gamma_{ml}$ represent the vertex from lattice simulations, the vertex from the results of SML, and the difference between the two results.}
\label{fig:vertex4}
\end{figure}
The renormalized Yukawa vertex is the simplest portal between the scalar singlet and the $SU(2)$ preserving Higgs fields in the form of condensate whose operator structure has an interesting role in the model's vacuum structure, beside being a combination of ultraviolet (UV) structure and infrared (IR) sensitive field propagators in the model. Being a higher correlation function than the propagators, it exhibits stronger fluctuations \cite{Maas:2013aia} against the same statistics, necessitating application of machine learning to develop an educated picture of the interaction. The function used for training is give by
\begin{equation}
f_{cost,\Gamma} = \sum_{m_{s},\alpha,\lambda,\beta,p,q} [\ F_{\Gamma}(m_{s},\alpha,\lambda,\beta,u,v) - \Gamma_{LS}(m_{s},\alpha,\lambda,\beta,p,q) ]\
\label{eq:costfuncvtx1}
\end{equation} 
where $u$ and $v$ have the same meaning as in \ref{eq:mlfuncvtx1}. The evolution of the cost function during training is shown in Figure \ref{fig:vtxcostfunc1} while the resulting coefficients are given in Appendix D. The dressing of the trained vertex is shown in Figures \ref{fig:vertex1}-\ref{fig:vertex4}, along with deviations between ML and LS results. As shown in the Figure \ref{fig:vtxcostfunc1}, the training proceeds well beyond the elbow point at which maximum convergence rate has occured effectively utilizing most of the capacity of the function being trained.
\par
The vertex in the form of dressing function is shown in Figures \ref{fig:vertex1}-\ref{fig:vertex4}. Despite fluctuations, qualitatively similar behavior of the vertex at all considered points in the parameter space is a remarkable feature in the model. The machine learnt vertex concurs with the lattice results despite possible presence of lattice artifacts in the Monte Carlo results and the limitations in the training of the scalar Yukawa vertex. The vertex is also found immune to the presence of $\phi^{4}$ interaction inside the model as well as other interaction couplings in the model, see Appendix $E$. It suggests a stable qualitative character of the vertex in the model. Furthermore, the trained dressing function of the vertex is not constant which is a strong indicator of contributions beyond tree level structure.
\par
A central observation is momentum dependence of the vertex dressing function. The decrease toward low field momenta indicates suppression of the off-shell vertex function relative to the subtraction scale. This behavior can be understood as a consequence of the amputation procedure. Since the propagators are infrared sensitive, a significant portion of the infrared effects in the vertex is removed upon amputation. Consequently, the vertex becomes primarily sensitive to the balance between interaction structure and the effective scaling regime of the lattice theory rather than directly reflecting strong infrared sensitivity as is the case with the field propagators. At this point, it is worth appreciating the role of the lattice scale, which is provided in appendix E, in the parameter space. Hence, the intermediate region of field momenta reveals the vertex interpolating between IR collective dynamics and UV sensitive lattice scaling structure, which is somewhat noticeable in the ML results.
\begin{figure} [h]
\centering
\includegraphics[width = 1.0\textwidth]{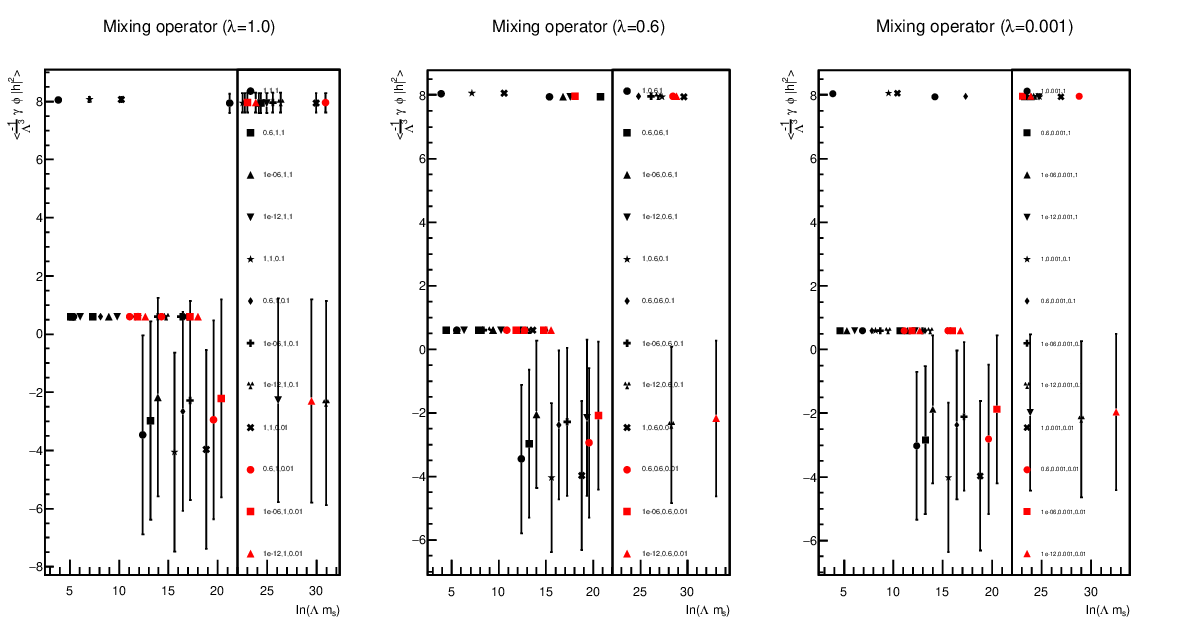}
\caption{Mixing operator derived from stationary structure against $(\alpha,\lambda,\gamma)$ is shown at different strengths of the Yukawa coupling.}
\label{fig:YukawaOperator1}
\end{figure}
\par
To further probe the lattice vertex relative to scaling, an operator $O_{Yuk}$ defined as $O_{Yuk}=<-\frac{1}{\Lambda^{3}} \gamma \phi h^{\dagger} h>$, with expectation performed over the entire Euclidean spacetime, is considered. Beside similarity to the vertex, a strong merit of the operator is that the stationary structure against the real singlet scalar field connects it to a mix of two smaller operators, which posses reduced statistical fluctuations. The expression is given by
\begin{equation}
<-\frac{1}{\Lambda^{3}}  \gamma \phi h^{\dagger} h> = \frac{1}{2 \Lambda^{3}} [m^{2}_{s} <\phi> + \lambda <h^{\dagger}h>],
\label{eq:stationaryWRTs1}
\end{equation} 
in which the two smaller correlation functions $<\phi>$ and $<h^{\dagger}h>$ mix, while ignoring the cubic self interaction of the real scalar field. Unambiguous existence of two branches is closely connected to the nontrivial scaling behavior of the lattice theory, see Figure \ref{fig:YukawaOperator1}. We observe that the effective scaling window mostly depends upon the quartic interaction couplings $\alpha$ and $\gamma$. Intermediate (lattice) mass values produce the broadest continuum-like scaling region, whereas both large and very small masses reduce the accessible scaling range due to discretization effects and infrared saturation, respectively. The quartic singlet self-coupling further modifies this behavior by controlling the stiffness of the singlet scalar fluctuations until the coupling is significantly weaker than $10^{-6}$, see Appendix E.
\begin{figure} [h]
\centering
\includegraphics[width = 1.0\textwidth]{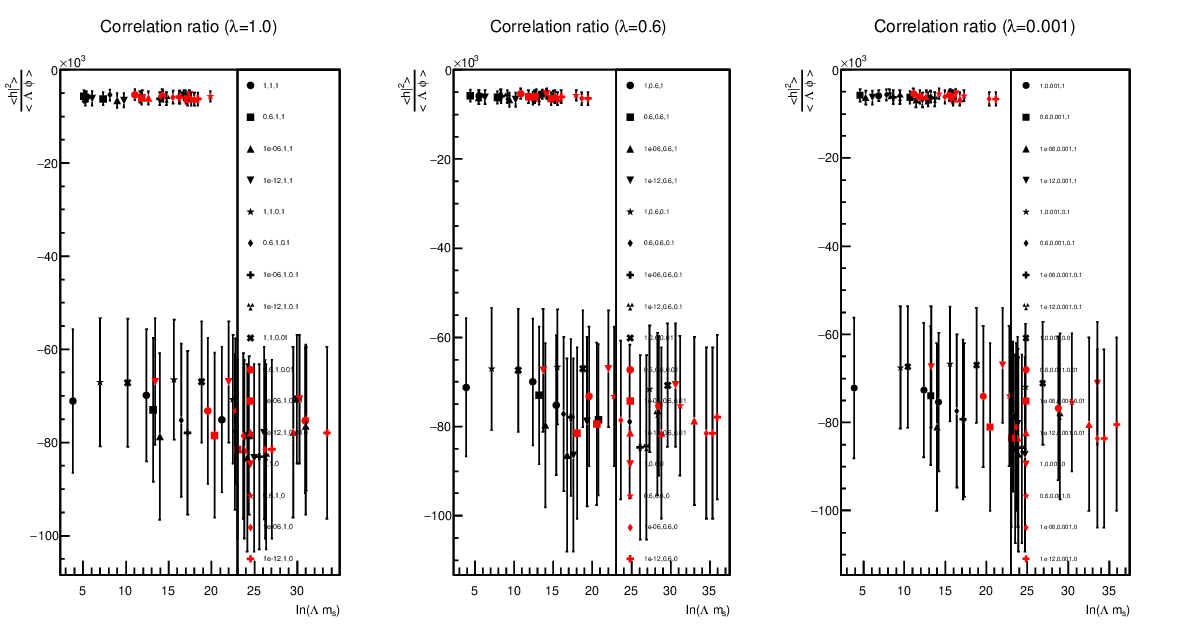}
\caption{Scalar expectation against $\Lambda m_{s}$ in GeVs at different strengths of Yukawa coupling $\lambda$ against the parameters $(\alpha,\lambda,\gamma)$ is shown.}
\label{fig:expectationratio1}
\end{figure}
The relation \label{eq:stationaryWRTs1} also serves as a demonstrator of a competition between the expectations $<\phi>$ and $<h^{\dagger}h>$, shown in Figure \ref{fig:expectationratio1} with respect to the scale. The distinction in the symmetry of the two operators is unambiguous as the singlet scalar expectation may also develop below the null value while the Higgs boson are prohibited, while the two fields are self-consistently mixed by the stationary condition \ref{eq:stationaryWRTs1}. The Figure \ref{fig:expectationratio1} supports stable proportionality between the two (singlet and doublet scalar) sectors in the presence of approximately constant regimes. Hence, we are observing a collective role of the singlet scalar and the portal to the $SU(2)$ sector in the form of stationary relationship, which is optimum of the potential with respect to the singlet scalar, in the form of two UV-driven regimes.
\section{Conclusion} \label{section:Conclusion}
The study of the field propagators, amputated Yukawa vertex, mixed composite operators, and scalar expectation values across a broad range of (lattice) bare parameters reveals a rich picture of the scalar sector. A central observation is the emergence of enhanced infrared behavior in the propagators together with dynamically generated negative effective mass-squared parameters extracted from low momentum propagator fits. At the same time, the computed scales remain organized and non-random across the parameter space. This indicates that the observed propagator enhancement is associated with a structured nonperturbative reorganization of the scalar sector rather than uncontrolled numerical instability.
\par
However, the amputated Yukawa vertex remains relatively stable and exhibits weak direct sensitivity to infrared scales. Amputation by the infrared enhanced propagators leaves behind a residual interaction kernel whose dominant structure remains comparatively stable due to the nonperturbative organization of correlations. The vertex serves as an elegant demonstrator of an interplay between the lattice scale and infrared physics.
\par
The mixed composite structure $<- \frac{1}{\Lambda^{3}} \phi h^{\dagger}h>$ computed from stationary condition reveals logarithmic branch structures against logarithmic bare scalar mass. The observed approximate mass insensitivity indicates that the dynamics of the sector is not controlled solely by perturbative mass scales. The qualitative behavior of the operator further indicates a strong dependence of the effective scaling window in the explored parameter region.
\par
The model exhibits a strong dependence of the effective continuum-like scaling window on both the bare singlet mass relative to the Higgs mass and the real singlet quartic self-coupling. The broadest scaling region is obtained for intermediate scalar masses, approximately two orders of magnitude smaller than the Higgs mass, while the quartic self-coupling further modifies this behavior by controlling the stiffness of scalar field fluctuations.
\par
The behavior of the model is consistent with an interplay between infrared enhanced propagation, dynamically reorganized scalar backgrounds, and ultraviolet sensitive scaling structure. Although, the present analysis does not establish thermodynamic criticality or continuum phase structure, it demonstrates that the scalar sector formed by a real singlet and an $SU(2)$ preserving complex doublet field exhibits stable nonperturbative correlation regimes simultaneously visible in propagators, interaction vertices, mixed composite operators, and scalar expectation values.
\section*{Acknowledgments}
We are deeply indebted to Lahore University of Management Sciences (LUMS) for providing their High Performance Computation facility to accomplish data collection, particularly Mr. Nouman Zubair for his efforts in ensuring a lubricant support of the facility, and generously funding the study. We are also indebted to Dr. Faisal Akram for valuable discussions regarding the investigation. A special Acknowledgement to Prof. Axel Maas who instilled interest in non-perturbative approaches in the first place.
\section*{Authors' Contributions}
\textit{Tajdar Mufti} is the principal investigator with intellectual contributions including conceiving of the research problem, development of algorithms for lattice simulations and libraries for analysis, and developing the manuscript. Collection of most of the statistics, arranging funds and computation resources are also included in his contributions to the investigation.\newline
\textit{Mohammad Saad} takes the credit of developing algorithms for lattice simulations and computations to verify the results and improving efficiency of libraries as a PhD candidate. He also contributed to pilot studies in the model before a full fledged data collection is started, and to analysis of the data from lattice simulations.
\section{Appendices}
\subsection{Appendix A: Scale ML coefficients}
\begin{table}[H]
    \centering
    \begin{tabular}{l|l|l|l|l}
    \hline
        i & j & k & l & ML coefficient \\ \hline
        0 & 0 & 0 & 0 & 0.00004204671611603871 \\
        1 & 0 & 0 & 0 & 0.00000033121372702243 \\
        0 & 1 & 0 & 0 & -0.00001699652709154267 \\
        0 & 0 & 1 & 0 & -0.00000624851907522656 \\
        0 & 0 & 0 & 1 & -0.00000341967311374059 \\
        2 & 0 & 0 & 0 & 0.00000019266241496355 \\
        1 & 1 & 0 & 0 & 0.00000065404354275954 \\
        1 & 0 & 1 & 0 & -0.00000254505528163963 \\
        1 & 0 & 0 & 1 & -0.00000014640152518038 \\
        0 & 2 & 0 & 0 & 0.00002137590007225947 \\ 
        0 & 1 & 1 & 0 & -0.00000166051840588897 \\ 
        0 & 1 & 0 & 1 & 0.00000007471965696236 \\ 
        0 & 0 & 2 & 0 & -0.00000307896125048292 \\ 
        0 & 0 & 1 & 1 & 0.00000523194323252075 \\ 
        0 & 0 & 0 & 2 & -0.00000884367220125723 \\ 
        3 & 0 & 0 & 0 & -0.00000000860134941660 \\ 
        2 & 1 & 0 & 0 & 0.00000024486255053333 \\ 
        2 & 0 & 1 & 0 & 0.00000004421188549401 \\ 
        2 & 0 & 0 & 1 & -0.00000003454791913580 \\ 
        1 & 2 & 0 & 0 & -0.00000254472272458965 \\ 
        1 & 1 & 1 & 0 & 0.00000204860138736421 \\ 
        1 & 1 & 0 & 1 & 0.00000073963873465633 \\ 
        1 & 0 & 2 & 0 & -0.00000037889798613823 \\ 
        1 & 0 & 1 & 1 & 0.00000067119886296475 \\ 
        1 & 0 & 0 & 2 & -0.00000041831175342321 \\
        0 & 3 & 0 & 0 & -0.00001231059343576337 \\ 
        0 & 2 & 1 & 0 & 0.00000982458892324012 \\ 
        0 & 2 & 0 & 1 & 0.00001021932299757391 \\ 
        0 & 1 & 2 & 0 & 0.00000349792886181747 \\ 
        0 & 1 & 1 & 1 & -0.00000114393730369691 \\
        0 & 1 & 0 & 2 & 0.00000038202005262826 \\ 
        0 & 0 & 3 & 0 & -0.00000238615232422338 \\ 
        0 & 0 & 2 & 1 & 0.00000920558955956622 \\ 
        0 & 0 & 1 & 2 & -0.00000329703262017962 \\
        0 & 0 & 0 & 3 & 0.00000177207291505899 \\ 
        4 & 0 & 0 & 0 & -0.00000000071975646433 \\
        3 & 1 & 0 & 0 & 0.00000000935893768757 \\ 
        3 & 0 & 1 & 0 & 0.00000000005266718655 \\ 
        3 & 0 & 0 & 1 & -0.00000000398654343259 \\
    \end{tabular}
\end{table}
\begin{table}[H]
    \centering
    \begin{tabular}{l|l|l|l|l}
    \hline
        i & j & k & l & ML coefficient \\ \hline
        2 & 2 & 0 & 0 & -0.00000006112840221978 \\ 
        2 & 1 & 1 & 0 & 0.00000004548057187473 \\ 
        2 & 1 & 0 & 1 & 0.00000007768391866151 \\ 
        2 & 0 & 2 & 0 & 0.00000001133305648615 \\ 
        2 & 0 & 1 & 1 & 0.00000001098803014700 \\ 
        2 & 0 & 0 & 2 & 0.00000001665589882381 \\ 
        1 & 3 & 0 & 0 & -0.00000005514379854280 \\ 
        1 & 2 & 1 & 0 & -0.00000028523729429545 \\ 
        1 & 2 & 0 & 1 & -0.00000108434450726203 \\ 
        1 & 1 & 2 & 0 & 0.00000027292969270976 \\ 
        1 & 1 & 1 & 1 & -0.00000082100061172858 \\ 
        1 & 1 & 0 & 2 & 0.00000090587885502577 \\ 
        1 & 0 & 3 & 0 & -0.00000003344214271414 \\ 
        1 & 0 & 2 & 1 & 0.00000003275645818923 \\ 
        1 & 0 & 1 & 2 & 0.00000000884265908950 \\ 
        1 & 0 & 0 & 3 & 0.00000002739669331155 \\ 
        0 & 4 & 0 & 0 & -0.00000346654097296598 \\ 
        0 & 3 & 1 & 0 & -0.00000108491311490566 \\ 
        0 & 3 & 0 & 1 & -0.00000148293683626711 \\ 
        0 & 2 & 2 & 0 & 0.00000107912366713592 \\ 
        0 & 2 & 1 & 1 & -0.00000451763776417841 \\
        0 & 2 & 0 & 2 & 0.00000325926638019423 \\ 
        0 & 1 & 3 & 0 & 0.00000280004248977691 \\ 
        0 & 1 & 2 & 1 & -0.00000441221362033836 \\ 
        0 & 1 & 1 & 2 & 0.00000099941279106748 \\ 
        0 & 1 & 0 & 3 & 0.00000014073561254104 \\ 
        0 & 0 & 4 & 0 & 0.00000001558082758722 \\ 
        0 & 0 & 3 & 1 & 0.00000168307077529126 \\ 
        0 & 0 & 2 & 2 & -0.00000024994842225213 \\ 
        0 & 0 & 1 & 3 & 0.00000060710259291389 \\ 
        0 & 0 & 0 & 4 & -0.00000056373519371439 \\
        5 & 0 & 0 & 0 & 0.00000000000618002916 \\ 
        4 & 1 & 0 & 0 & 0.00000000037421839958 \\ 
        4 & 0 & 1 & 0 & -0.00000000001582683863 \\ 
        4 & 0 & 0 & 1 & -0.00000000012539899164 \\ 
        3 & 2 & 0 & 0 & 0.00000000252715908005 \\ 
        3 & 1 & 1 & 0 & -0.00000000346498659530 \\ 
        3 & 1 & 0 & 1 & 0.00000000003667919136 \\ 
        3 & 0 & 2 & 0 & 0.00000000438372584575 \\ 
    \end{tabular}
\end{table}
\begin{table}[H]
    \centering
    \begin{tabular}{l|l|l|l|l}
    \hline
        i & j & k & l & ML coefficient \\ \hline
        3 & 0 & 1 & 1 & -0.00000000040836259939 \\ 
        3 & 0 & 0 & 2 & -0.00000000102358138047 \\ 
        2 & 3 & 0 & 0 & -0.00000002320830031735 \\ 
        2 & 2 & 1 & 0 & -0.00000000394795812741 \\ 
        2 & 2 & 0 & 1 & -0.00000002626441722203 \\ 
        2 & 0 & 3 & 0 & 0.00000002507747562543 \\ 
        2 & 0 & 2 & 1 & -0.00000000294643202813 \\
        2 & 0 & 1 & 2 & -0.00000000563436871654 \\ 
        2 & 0 & 0 & 3 & 0.00000001384677046022 \\ 
        1 & 4 & 0 & 0 & -0.00000006895340961140 \\ 
        1 & 3 & 1 & 0 & -0.00000042097215880076 \\ 
        1 & 3 & 0 & 1 & -0.00000005012822378196 \\ 
        1 & 2 & 2 & 0 & -0.00000013791880996462 \\ 
        1 & 2 & 1 & 1 & -0.00000001854045829703 \\ 
        1 & 2 & 0 & 2 & 0.00000018203980725287 \\ 
        1 & 1 & 3 & 0 & 0.00000009979780106344 \\ 
        1 & 1 & 2 & 1 & -0.00000066879624643264 \\ 
        1 & 1 & 1 & 2 & 0.00000054541212646682 \\ 
        1 & 1 & 0 & 3 & -0.00000011435625299090 \\
        1 & 0 & 4 & 0 & -0.00000002823884136080 \\ 
        1 & 0 & 3 & 1 & 0.00000002818435245999 \\ 
        1 & 0 & 2 & 2 & 0.00000010113190090918 \\ 
        1 & 0 & 1 & 3 & -0.00000022882713791289 \\
        1 & 0 & 0 & 4 & -0.00000004590930772402 \\ 
        0 & 5 & 0 & 0 & -0.00000100314542479732 \\ 
        0 & 4 & 1 & 0 & -0.00000048971119205572 \\ 
        0 & 4 & 0 & 1 & -0.00000074462390219540 \\ 
        0 & 3 & 2 & 0 & -0.00000155136758717685 \\ 
        0 & 3 & 1 & 1 & -0.00000223714866767327 \\ 
        0 & 3 & 0 & 2 & 0.00000082571644575295 \\ 
        0 & 2 & 3 & 0 & 0.00000204973802199166 \\ 
        0 & 2 & 2 & 1 & -0.00000409177470849451 \\ 
        0 & 2 & 1 & 2 & 0.00000363027896509118 \\ 
        0 & 2 & 0 & 3 & -0.00000033485339250591 \\ 
        0 & 1 & 4 & 0 & 0.00000177563501443542 \\ 
        0 & 1 & 3 & 1 & -0.00000255896770830752 \\ 
        0 & 1 & 2 & 2 & 0.00000000101899899756 \\ 
        0 & 1 & 1 & 3 & -0.00000026016168342555 \\
        0 & 1 & 0 & 4 & 0.00000093511088621316 \\ 
    \end{tabular}
\end{table}
\begin{table}[H]
    \centering
    \begin{tabular}{l|l|l|l|l}
    \hline
        i & j & k & l & ML coefficient \\ \hline
        0 & 0 & 5 & 0 & -0.00000088614285353744 \\ 
        0 & 0 & 4 & 1 & -0.00000002235260200843 \\ 
        0 & 0 & 3 & 2 & 0.00000131788233937997 \\ 
        0 & 0 & 2 & 3 & -0.00000035632870144094 \\ 
        0 & 0 & 1 & 4 & 0.00000050901791952562 \\ 
        0 & 0 & 0 & 5 & -0.00000042881150120108 \\ 
    \end{tabular}
\end{table}
\subsection{Appendix B: Scalar propagator ML coefficients}
\begin{table}[H]
    \centering
    \begin{tabular}{l|l|l|l|l|l}
    \hline
        i & j & k & l & m & ML coefficient \\ \hline
        0 & 0 & 0 & 0 & 0 & -29249117.68976046338320884388 \\ 
        1 & 0 & 0 & 0 & 0 & 4494304.00959605802563601173 \\ 
        0 & 1 & 0 & 0 & 0 & 5051541300.10930459434166550636 \\
        0 & 0 & 1 & 0 & 0 & -3059653.42608087466123834020 \\ 
        0 & 0 & 0 & 1 & 0 & -712387003.12874731619376689196 \\ 
        0 & 0 & 0 & 0 & 1 & 243593.81333398239533494234 \\ 
        2 & 0 & 0 & 0 & 0 & 4031883.48816977344631595770 \\ 
        1 & 1 & 0 & 0 & 0 & 9826134618.49047317076474428177 \\
        1 & 0 & 1 & 0 & 0 & -560233.60726212905768761630 \\ 
        1 & 0 & 0 & 1 & 0 & -15830315485.36271013971418142319 \\ 
        1 & 0 & 0 & 0 & 1 & 469599.40675439177286421000 \\ 
        0 & 2 & 0 & 0 & 0 & 32655255176.60140623711049556732 \\
        0 & 1 & 1 & 0 & 0 & 516653.29528092876148548385 \\ 
        0 & 1 & 0 & 1 & 0 & -4470185557020.13627004623413085938 \\
        0 & 1 & 0 & 0 & 1 & -587221233.59788914368255063891 \\
        0 & 0 & 2 & 0 & 0 & -510287.42027255551312237003 \\ 
        0 & 0 & 1 & 1 & 0 & -4605002490.18229327024891972542 \\ 
        0 & 0 & 1 & 0 & 1 & 1619103.70655767135747282737 \\ 
        0 & 0 & 0 & 2 & 0 & 61302310914629.28273391723632812500 \\ 
        0 & 0 & 0 & 1 & 1 & -18270534615.35174795798957347870 \\
        0 & 0 & 0 & 0 & 2 & -899087.44560125185074639376 \\ 
        3 & 0 & 0 & 0 & 0 & 311233805.69143117495696060359 \\ 
        2 & 1 & 0 & 0 & 0 & 202370833119.46218262612819671631 \\ 
        2 & 0 & 1 & 0 & 0 & -8079971.98481350576412296505 \\ 
        2 & 0 & 0 & 1 & 0 & -181910749170.29909703135490417480 \\ 
        2 & 0 & 0 & 0 & 1 & 136782687.54878008981177117676 \\ 
        1 & 2 & 0 & 0 & 0 & -18242045254447.52832221984863281250 \\ 
        1 & 1 & 1 & 0 & 0 & 3114338091.76502288808114826679 \\ 
        1 & 1 & 0 & 1 & 0 & -439765012379.60115143656730651855 \\ 
        1 & 1 & 0 & 0 & 1 & -5896145300.85646927729249000549 \\ 
        1 & 0 & 2 & 0 & 0 & 1412694.44728011203244477656 \\ 
        1 & 0 & 1 & 1 & 0 & -28857767829.40845576673746109009 \\ 
        1 & 0 & 1 & 0 & 1 & -774907.92920788870952719662 \\
        1 & 0 & 0 & 2 & 0 & 537905231022383.49450683593750000000 \\
        1 & 0 & 0 & 1 & 1 & -15571540126.14118602219969034195 \\ 
        1 & 0 & 0 & 0 & 2 & -15418521.05064824574765225407 \\
        0 & 3 & 0 & 0 & 0 & -3818340990943901.39306640625 \\ 
        0 & 2 & 1 & 0 & 0 & -2929159678550.48598337173461914062 \\ 
        0 & 2 & 0 & 1 & 0 & 5304301023144441.072265625 \\
    \end{tabular}
\end{table}
\begin{table}[H]
    \centering
    \begin{tabular}{l|l|l|l|l|l}
    \hline
        i & j & k & l & m & ML coefficient \\ \hline
        0 & 2 & 0 & 0 & 1 & -4136880472392.41243124008178710938 \\
        0 & 1 & 2 & 0 & 0 & 1154958248.25882628152612596750 \\ 
        0 & 1 & 1 & 1 & 0 & 27775336037476.92648315429687500000 \\ 
        0 & 1 & 1 & 0 & 1 & 4841300576.08393410639837384224 \\
        0 & 1 & 0 & 2 & 0 & 37999004020547656.19921875000000000000 \\
        0 & 1 & 0 & 1 & 1 & 910564724235.55115342140197753906 \\ 
        0 & 1 & 0 & 0 & 2 & -5881121.14464149881405319320 \\ 
        0 & 0 & 3 & 0 & 0 & -399049.17893793460774531923 \\ 
        0 & 0 & 2 & 1 & 0 & -13882693351.84844592120498418808 \\ 
        0 & 0 & 2 & 0 & 1 & -306043.06604617952851299378 \\ 
        0 & 0 & 1 & 2 & 0 & -1036220733052.04271733760833740234 \\
        0 & 0 & 1 & 1 & 1 & 5679029253.67762702563777565956 \\
        0 & 0 & 1 & 0 & 2 & 88627.84113343548440866471 \\ 
        0 & 0 & 0 & 3 & 0 & 6188276259827914650.5 \\ 
        0 & 0 & 0 & 2 & 1 & -86628963937697.49748992919921875000 \\ 
        0 & 0 & 0 & 1 & 2 & 281162.78052788321014077155 \\ 
        0 & 0 & 0 & 0 & 3 & -2655756.52361690355724022083 \\ 
        4 & 0 & 0 & 0 & 0 & 34671765.04011332819572999142 \\ 
        3 & 1 & 0 & 0 & 0 & -45688006299.24168493971228599548 \\ 
        3 & 0 & 1 & 0 & 0 & 56874057.19820563030225457624 \\ 
        3 & 0 & 0 & 1 & 0 & -816629138194.41110175848007202148 \\ 
        3 & 0 & 0 & 0 & 1 & -26716326.78837980910975602455 \\
        2 & 2 & 0 & 0 & 0 & 6458619708152.00710582733154296875 \\
        2 & 1 & 1 & 0 & 0 & 34079264550.88940015248954296112 \\ 
        2 & 1 & 0 & 1 & 0 & -463041765350783.988555908203125 \\ 
        2 & 1 & 0 & 0 & 1 & -2278057665.31500764470547437668 \\
        2 & 0 & 2 & 0 & 0 & 56990382.04602564356900984421 \\ 
        2 & 0 & 1 & 1 & 0 & -742273992515.96871608495712280273 \\ 
        2 & 0 & 1 & 0 & 1 & 939049.40311035307320253196 \\
        2 & 0 & 0 & 2 & 0 & 6113211471729463.4775390625 \\ 
        2 & 0 & 0 & 2 & 2 & -97421822050.64413222670555114746 \\ 
        2 & 0 & 0 & 0 & 2 & 5351318.61784618518959177891 \\
        1 & 3 & 0 & 0 & 0 & -18025666159318444.634765625 \\ 
        1 & 2 & 1 & 0 & 0 & -8373496453936.19244527816772460938 \\ 
        1 & 2 & 0 & 1 & 0 & 12950287249157703.771484375 \\ 
        1 & 2 & 0 & 0 & 1 & -4002954677613.27053308486938476562 \\ 
        1 & 1 & 2 & 0 & 0 & 2481318913.91772613185457885265 \\
        1 & 1 & 1 & 1 & 0 & 151698596523870.50750732421875 \\ 
        1 & 1 & 1 & 0 & 1 & -2824801983.60942707234062254429 \\
    \end{tabular}
\end{table}
\begin{table}[H]
    \centering
    \begin{tabular}{l|l|l|l|l|l}
    \hline
        i & j & k & l & m & ML coefficient \\ \hline
        1 & 1 & 0 & 2 & 0 & 324850199623930749.96875000000000000000 \\
        1 & 1 & 0 & 1 & 1 & 16704423414240.96300601959228515625 \\ 
        1 & 1 & 0 & 0 & 2 & 592612053.85543076822068542242 \\ 
        1 & 0 & 3 & 0 & 0 & 4515315.30390288060516468249 \\ 
        1 & 0 & 2 & 1 & 0 & -89587736429.20954307168722152710 \\ 
        1 & 0 & 2 & 0 & 1 & 1084932.79894708410620296490 \\
        1 & 0 & 1 & 2 & 0 & 299215565290503.9815673828125 \\ 
        1 & 0 & 1 & 1 & 1 & -29286126104.04314007796347141266 \\ 
        1 & 0 & 1 & 0 & 2 & 755064.02451749223320121018 \\
        1 & 0 & 0 & 3 & 0 & 11312546829116742981.0 \\ 
        1 & 0 & 0 & 2 & 1 & -39801554601089.67571640014648437500 \\ 
        1 & 0 & 0 & 1 & 2 & 22604754698.61873258650302886963 \\
        1 & 0 & 0 & 0 & 3 & -222844.90253171941731125116 \\ 
        0 & 4 & 0 & 0 & 0 & -847200704373069437.3125 \\ 
        0 & 3 & 1 & 0 & 0 & -3697237474966760.9873046875 \\ 
        0 & 3 & 0 & 1 & 0 & -4668647372767127579.0 \\ 
        0 & 3 & 0 & 0 & 1 & -1807361196570560.34094238281250000000 \\ 
        0 & 2 & 2 & 0 & 0 & -1325893707728.79177522659301757812 \\ 
        0 & 2 & 1 & 1 & 0 & 49077783129159335.76953125 \\ 
        0 & 2 & 1 & 0 & 1 & -2508999002067.24148368835449218750 \\ 
        0 & 2 & 0 & 2 & 0 & 100.0 \\
        0 & 2 & 0 & 1 & 1 & -1866098216015847.926513671875 \\ 
        0 & 2 & 0 & 0 & 2 & -38338369948.64360289275646209717 \\ 
        0 & 1 & 3 & 0 & 0 & 21443.94461924049635470624 \\
        0 & 1 & 2 & 1 & 0 & 48324143836907.33441925048828125 \\ 
        0 & 1 & 2 & 0 & 1 & -1856852071.56813417188823223114 \\
        0 & 1 & 1 & 2 & 0 & 153832919495623059.54687500000000000000 \\
        0 & 1 & 1 & 1 & 1 & -688587073266.42043524980545043945 \\
        0 & 1 & 1 & 0 & 2 & 473629849.17630420209025032818 \\
        0 & 1 & 0 & 3 & 0 & 0.00000000000000000000 \\
        0 & 1 & 0 & 2 & 1 & -180463019615703665.10937500000000000000 \\ 
        0 & 1 & 0 & 1 & 2 & 7742931519199.54397773742675781250 \\
        0 & 1 & 0 & 0 & 3 & -89906073.82410915187938371673 \\ 
        0 & 0 & 4 & 0 & 0 & -1702961.87954536448808084970 \\ 
        0 & 0 & 3 & 1 & 0 & -31683445082.24457602947950363159 \\ 
        0 & 0 & 3 & 0 & 1 & 46651.58360522819130977723 \\ 
        0 & 0 & 2 & 2 & 0 & 19209798837851.55890655517578125000 \\
        0 & 0 & 2 & 1 & 1 & -9962293132.64522106572985649109 \\ 
        0 & 0 & 2 & 0 & 2 & -24907.21562488712053173856 \\
    \end{tabular}
\end{table}
\subsection{Appendix C: Higgs propagator ML coefficients}
\begin{table}[H]
    \centering
    \begin{tabular}{l|l|l|l|l|l}
    \hline
        i & j & k & l & m & ML coefficient \\ \hline
        0 & 0 & 1 & 3 & 0 & 3826591519676916228.50000000000000000000 \\
        0 & 0 & 1 & 2 & 1 & 102454445834943.10730743408203125000 \\ 
        0 & 0 & 1 & 1 & 2 & 4317786428.19358752435073256493 \\ 
        0 & 0 & 1 & 0 & 3 & 131775.21727110358713730420 \\ 
        0 & 0 & 0 & 4 & 0 & 0.00000000000000000000 \\ 
        0 & 0 & 0 & 3 & 1 & 104914021462.90689944475889205933 \\ 
        0 & 0 & 0 & 2 & 2 & 12036200169856.03668880462646484375 \\
        0 & 0 & 0 & 1 & 3 & 5298887355.80036131944507360458 \\ 
        0 & 0 & 0 & 0 & 4 & -2394787.46599786293541001214 \\ 
        0 & 0 & 0 & 0 & 0 & -445453912.47500000000582076609 \\
        1 & 0 & 0 & 0 & 0 & 407523128.61749999999301508069 \\ 
        0 & 1 & 0 & 0 & 0 & 4988800000.00000000000000000000 \\
        0 & 0 & 1 & 0 & 0 & -177933423.97499999999126885086 \\
        0 & 0 & 0 & 1 & 0 & 4967000000.00000000000000000000 \\
        0 & 0 & 0 & 0 & 1 & 58563894.16249999999490682967 \\ 
        2 & 0 & 0 & 0 & 0 & 2600559159.62500000000000000000 \\ 
        1 & 1 & 0 & 0 & 0 & 4990800000.00000000000000000000 \\ 
        1 & 0 & 1 & 0 & 0 & -243290430.32499999999708961695 \\ 
        1 & 0 & 0 & 1 & 0 & 4970800000.00000000000000000000 \\ 
        1 & 0 & 0 & 0 & 1 & -151756075.86249999998835846782 \\ 
        0 & 2 & 0 & 0 & 0 & 4988600000.00000000000000000000 \\ 
        0 & 1 & 1 & 0 & 0 & 4990172500.00000000000000000000 \\ 
        0 & 1 & 0 & 1 & 0 & 4990400000.00000000000000000000 \\ 
        0 & 1 & 0 & 0 & 1 & 1658444612.50000000000000000000 \\ 
        0 & 0 & 2 & 0 & 0 & 240383721.98950000001059379429 \\ 
        0 & 0 & 1 & 1 & 0 & 19687227.50000000000000000000 \\ 
        0 & 0 & 1 & 0 & 1 & 26855384.99023750000378640834 \\ 
        0 & 0 & 0 & 2 & 0 & 4971400000.00000000000000000000 \\ 
        0 & 0 & 0 & 1 & 1 & 850204510.00000000000000000000 \\ 
        0 & 0 & 0 & 0 & 2 & -20171151.09966250000616128091 \\ 
        3 & 0 & 0 & 0 & 0 & 158466932.16749999998137354851 \\ 
        2 & 1 & 0 & 0 & 0 & 4997400000.00000000000000000000 \\ 
        2 & 0 & 1 & 0 & 0 & 2427111504.74875000002793967724 \\ 
        2 & 0 & 0 & 1 & 0 & 4991400000.00000000000000000000 \\ 
        2 & 0 & 0 & 0 & 1 & 955890464.00000000000000000000 \\ 
        1 & 2 & 0 & 0 & 0 & 4991800000.00000000000000000000 \\ 
        1 & 1 & 1 & 0 & 0 & 4291177112.50000000000000000000 \\ 
        1 & 1 & 0 & 1 & 0 & 4992400000.00000000000000000000 \\ 
        1 & 1 & 0 & 0 & 1 & 569658552.50000000000000000000 \\ 
    \end{tabular}
\end{table}
\begin{table}[H]
    \centering
    \begin{tabular}{l|l|l|l|l|l}
    \hline
        i & j & k & l & m & ML coefficient \\ \hline
        1 & 0 & 2 & 0 & 0 & -83854561.36171250000916188583 \\ 
        1 & 0 & 1 & 1 & 0 & 2114090540.10000000009313225746 \\ 
        1 & 0 & 1 & 0 & 1 & 38274515.90726000001450302079 \\ 
        1 & 0 & 0 & 2 & 0 & 4976000000.00000000000000000000 \\ 
        1 & 0 & 0 & 1 & 1 & 2150570630.00000000000000000000 \\ 
        1 & 0 & 0 & 0 & 2 & 104171524.06086250003863824531 \\ 
        0 & 3 & 0 & 0 & 0 & 4988400000.00000000000000000000 \\ 
        0 & 2 & 1 & 0 & 0 & 4707387500.00000000000000000000 \\ 
        0 & 2 & 0 & 1 & 0 & 4990000000.00000000000000000000 \\ 
        0 & 2 & 0 & 0 & 1 & -1429866250.00000000000000000000 \\ 
        0 & 1 & 2 & 0 & 0 & 3822368511.75000000000000000000 \\ 
        0 & 1 & 1 & 1 & 0 & 2651585000.00000000000000000000 \\ 
        0 & 1 & 1 & 0 & 1 & -1531032500.00000000000000000000 \\ 
        0 & 1 & 0 & 2 & 0 & 4990000000.00000000000000000000 \\ 
    \end{tabular}
\end{table}
\begin{table}[H]
    \centering
    \begin{tabular}{l|l|l|l|l|l}
    \hline
        i & j & k & l & m & ML coefficient \\ \hline
        0 & 1 & 0 & 1 & 1 & -1732600000.00000000000000000000 \\ 
        0 & 1 & 0 & 0 & 2 & -1424923450.75000000000000000000 \\ 
        0 & 0 & 3 & 0 & 0 & -121547974.10103750000416766852 \\ 
        0 & 0 & 2 & 1 & 0 & 2017510509.25000000000000000000 \\ 
        0 & 0 & 2 & 0 & 1 & 35582230.44033499999932246283 \\ 
        0 & 0 & 1 & 2 & 0 & 1524917500.00000000000000000000 \\
        0 & 0 & 1 & 1 & 1 & 561043223.07500000001164153218 \\ 
        0 & 0 & 1 & 0 & 2 & 203672.84485950000031095897 \\ 
        0 & 0 & 0 & 3 & 0 & 4974600000.00000000000000000000 \\ 
        0 & 0 & 0 & 2 & 1 & 2625350000.00000000000000000000 \\ 
        0 & 0 & 0 & 1 & 2 & 2299144597.53750000009313225746 \\ 
        0 & 0 & 0 & 0 & 3 & -33540623.13344632486405316740 \\ 
        4 & 0 & 0 & 0 & 0 & -23272104.53164249999463208951 \\ 
        3 & 1 & 0 & 0 & 0 & -4468729087.75025000004097819328 \\ 
        3 & 0 & 1 & 0 & 0 & -29437088.56651250001959851943 \\ 
        3 & 0 & 0 & 1 & 0 & -4987000000.00000000000000000000 \\ 
        3 & 0 & 0 & 0 & 1 & -7867769.59300000000075669959 \\ 
        2 & 2 & 0 & 0 & 0 & 4997600000.00000000000000000000 \\ 
        2 & 1 & 1 & 0 & 0 & 4303736650.00000000000000000000 \\ 
        2 & 1 & 0 & 1 & 0 & 4997600000.00000000000000000000 \\ 
        2 & 1 & 0 & 0 & 1 & -1431317500.00000000000000000000 \\ 
        2 & 0 & 2 & 0 & 0 & 329994425.89750000002095475793 \\ 
        2 & 0 & 1 & 1 & 0 & 1302345100.00000000000000000000 \\ 
        2 & 0 & 1 & 0 & 1 & -871624992.04050000000279396772 \\ 
        2 & 0 & 0 & 2 & 0 & 4990200000.00000000000000000000 \\ 
        2 & 0 & 0 & 2 & 2 & 2413760297.50000000000000000000 \\ 
        2 & 0 & 0 & 0 & 2 & -123256292.28674999996292172000 \\ 
        1 & 3 & 0 & 0 & 0 & 4990000000.00000000000000000000 \\ 
        1 & 2 & 1 & 0 & 0 & 4682427575.00000000000000000000 \\ 
        1 & 2 & 0 & 1 & 0 & 4992200000.00000000000000000000 \\ 
        1 & 2 & 0 & 0 & 1 & -1239405000.00000000000000000000 \\
        1 & 1 & 2 & 0 & 0 & 4214006965.82500000018626451492 \\ 
        1 & 1 & 1 & 1 & 0 & 2862675000.00000000000000000000 \\ 
        1 & 1 & 1 & 0 & 1 & -923595325.00000000000000000000 \\ 
        1 & 1 & 0 & 2 & 0 & 4992200000.00000000000000000000 \\ 
        1 & 1 & 0 & 1 & 1 & -1483600000.00000000000000000000 \\ 
        1 & 1 & 0 & 0 & 2 & -1006049393.25000000000000000000 \\ 
        1 & 0 & 3 & 0 & 0 & 108432818.27578749992244411260 \\ 
        1 & 0 & 2 & 1 & 0 & 26877576.37500000000000000000 \\ 
    \end{tabular}
\end{table}
\begin{table}[H]
    \centering
    \begin{tabular}{l|l|l|l|l|l}
    \hline
        i & j & k & l & m & ML coefficient \\ \hline
        1 & 0 & 2 & 0 & 1 & -86865834.08642625003267312422 \\ 
        1 & 0 & 1 & 2 & 0 & 2919674995.00000000000000000000 \\ 
        1 & 0 & 1 & 1 & 1 & 2682795875.75000000000000000000 \\ 
        1 & 0 & 1 & 0 & 2 & 62505438.62932499997623381205 \\ 
        1 & 0 & 0 & 3 & 0 & 4977095000.00000000000000000000 \\ 
        1 & 0 & 0 & 2 & 1 & 4879025000.00000000000000000000 \\ 
        1 & 0 & 0 & 1 & 2 & 4236970237.60000000009313225746 \\ 
        1 & 0 & 0 & 0 & 3 & -31958740.70953999998891958967 \\ 
        0 & 4 & 0 & 0 & 0 & 4987000000.00000000000000000000 \\ 
        0 & 3 & 1 & 0 & 0 & 4857800000.00000000000000000000 \\ 
        0 & 3 & 0 & 1 & 0 & 4971628750.00000000000000000000 \\ 
        0 & 3 & 0 & 0 & 1 & -1313301250.00000000000000000000 \\ 
        0 & 2 & 2 & 0 & 0 & 4113528525.00000000000000000000 \\ 
        0 & 2 & 1 & 1 & 0 & 3090680000.00000000000000000000 \\ 
        0 & 2 & 1 & 0 & 1 & -977581250.00000000000000000000 \\ 
        0 & 2 & 0 & 2 & 0 & 3908858500.00000000000000000000 \\ 
        0 & 2 & 0 & 1 & 1 & -1572748775.00000000000000000000 \\ 
        0 & 2 & 0 & 0 & 2 & -1366220375.00000000000000000000 \\ 
        0 & 1 & 3 & 0 & 0 & 3801964652.25000000000000000000 \\ 
        0 & 1 & 2 & 1 & 0 & 2831498750.00000000000000000000 \\ 
        0 & 1 & 2 & 0 & 1 & -1498662500.00000000000000000000 \\ 
        0 & 1 & 1 & 2 & 0 & 2764856100.00000000000000000000 \\ 
        0 & 1 & 1 & 1 & 1 & -1122200000.00000000000000000000 \\
        0 & 1 & 1 & 0 & 2 & 718353902.25000000000000000000 \\ 
        0 & 1 & 0 & 3 & 0 & 1315315000.00000000000000000000 \\ 
        0 & 1 & 0 & 2 & 1 & -2399400000.00000000000000000000 \\ 
        0 & 1 & 0 & 1 & 2 & -1896170000.00000000000000000000 \\ 
        0 & 1 & 0 & 0 & 3 & -1112579090.57499999995343387127 \\
        0 & 0 & 4 & 0 & 0 & -71719301.53672499999083811417 \\ 
        0 & 0 & 3 & 1 & 0 & -361869483.12500000000000000000 \\ 
        0 & 0 & 3 & 0 & 1 & 11942533.88535275003232527524 \\ 
        0 & 0 & 2 & 2 & 0 & 2120483000.00000000000000000000 \\ 
        0 & 0 & 2 & 1 & 1 & 2120446234.59987499995622783899 \\ 
        0 & 0 & 2 & 0 & 2 & -1411265.23503450050327501231 \\ 
        0 & 0 & 1 & 3 & 0 & 1321045125.00000000000000000000 \\ 
        0 & 0 & 1 & 2 & 1 & -1646670117.50000000000000000000 \\
        0 & 0 & 1 & 1 & 2 & 69770627.06200000015087425709 \\ 
        0 & 0 & 1 & 0 & 3 & -10839710.28153034254683007021 \\
        0 & 0 & 0 & 4 & 0 & 49062500.00000000000000000000 \\
    \end{tabular}
\end{table}
\begin{table}[H]
    \centering
    \begin{tabular}{l|l|l|l|l|l}
    \hline
        i & j & k & l & m & ML coefficient \\ \hline
        0 & 0 & 0 & 3 & 1 & 4476977200.00000000000000000000 \\
        0 & 0 & 0 & 2 & 2 & 4937525000.00000000000000000000 \\
        0 & 0 & 0 & 1 & 3 & 3503393756.02499999990686774254 \\
        0 & 0 & 0 & 0 & 4 & -28293589.75789467481627070811 \\
    \end{tabular}
\end{table}
\subsection{Appendix D vertex ML coefficients}
\begin{table}[H]
    \centering
    \begin{tabular}{l|l|l|l|l|l|l}
    \hline
        i & j & k & l & m & n & ML coefficient \\ \hline
        0 & 0 & 0 & 0 & 0 & 0 & -3.45526331236478194604 \\ 
        1 & 0 & 0 & 0 & 0 & 0 & 4803.03634376115783322092 \\
        0 & 1 & 0 & 0 & 0 & 0 & -56.33389964582678480221 \\ 
        0 & 0 & 1 & 0 & 0 & 0 & 198570.97986139306132713500 \\ 
        0 & 0 & 0 & 1 & 0 & 0 & -312.61780115174700395153 \\ 
        0 & 0 & 0 & 0 & 1 & 0 & 2929.50181024512530658654 \\ 
        0 & 0 & 0 & 0 & 0 & 1 & 19.36617026549533105914 \\ 
        2 & 0 & 0 & 0 & 0 & 0 & -731346351.17602740437723696232 \\ 
        1 & 1 & 0 & 0 & 0 & 0 & 332199.96523956582873893240 \\ 
        1 & 0 & 1 & 0 & 0 & 0 & -469252018.33761127636535093188 \\ 
        1 & 0 & 0 & 1 & 0 & 0 & -398122.91294051367009387832 \\ 
        1 & 0 & 0 & 0 & 1 & 0 & 22881552.14189981575327692553 \\ 
        1 & 0 & 0 & 0 & 0 & 1 & 43071.47982171925457706152 \\ 
        0 & 2 & 0 & 0 & 0 & 0 & -337.64075158703259374460 \\ 
        0 & 1 & 1 & 0 & 0 & 0 & 3709790.40265101098952982284 \\ 
        0 & 1 & 0 & 1 & 0 & 0 & -435.57738885287709565408 \\ 
        0 & 1 & 0 & 0 & 1 & 0 & 39391.41888330093927450548 \\ 
        0 & 1 & 0 & 0 & 0 & 1 & 6.87642735411610318786 \\ 
        0 & 0 & 2 & 0 & 0 & 0 & -26951497419.61732320673763751984 \\ 
        0 & 0 & 1 & 1 & 0 & 0 & -445791.21378722193443877586 \\ 
        0 & 0 & 1 & 0 & 1 & 0 & 42619689.67281903805633191951 \\ 
        0 & 0 & 1 & 0 & 0 & 1 & 42054.92229840011366093222 \\ 
        0 & 0 & 0 & 2 & 0 & 0 & -319.06056866570313423903 \\ 
        0 & 0 & 0 & 1 & 1 & 0 & -64328.07704982001000715286 \\
        0 & 0 & 0 & 1 & 0 & 1 & 252.68425591544965695678 \\ 
        0 & 0 & 0 & 0 & 2 & 0 & 7231.57195752250196640887 \\ 
        0 & 0 & 0 & 0 & 1 & 1 & 10603.05472494199731148257 \\ 
        0 & 0 & 0 & 0 & 0 & 2 & -14.90556180742878812379 \\
        3 & 0 & 0 & 0 & 0 & 0 & 2063492380252.85643410682678222656 \\
        2 & 1 & 0 & 0 & 0 & 0 & -1329525770.38859826035331934690 \\ 
        2 & 0 & 1 & 0 & 0 & 0 & 3775911486447.34214520454406738281 \\
        2 & 0 & 0 & 1 & 0 & 0 & 577455417.31073282501893118024 \\ 
        2 & 0 & 0 & 0 & 1 & 0 & -466120993938.84940174221992492676 \\
        2 & 0 & 0 & 0 & 0 & 1 & 274281267.40107618217007257044 \\
        1 & 2 & 0 & 0 & 0 & 0 & 46882.58100104748680792000 \\
        1 & 1 & 1 & 0 & 0 & 0 & -5889236204.28922640252858400345 \\
        1 & 1 & 0 & 1 & 0 & 0 & 702006.91363271247161037536 \\ 
        1 & 1 & 0 & 0 & 1 & 0 & 57485801.03282510591088794172 \\
        1 & 1 & 0 & 0 & 0 & 1 & -169541.43650110202484881938 \\ 
    \end{tabular}
\end{table}
\begin{table}[H]
    \centering
    \begin{tabular}{l|l|l|l|l|l|l}
    \hline
        i & j & k & l & m & n & ML coefficient \\ \hline
        1 & 0 & 2 & 0 & 0 & 0 & -46651474364427.43116378784179687500 \\
        1 & 0 & 1 & 1 & 0 & 0 & -11971217460.84712700638920068741 \\ 
        1 & 0 & 1 & 0 & 1 & 0 & 547070436389.73784652352333068848 \\ 
        1 & 0 & 1 & 0 & 0 & 1 & 215182918.63060795450292062014 \\ 
        1 & 0 & 0 & 2 & 0 & 0 & -499758.62681993621009723938 \\ 
        1 & 0 & 0 & 1 & 1 & 0 & -75971545.67342204385931836441 \\ 
        1 & 0 & 0 & 1 & 0 & 1 & -54038.97801880424650633472 \\ 
        1 & 0 & 0 & 0 & 2 & 0 & 3511896.38563321286119389697 \\ 
        1 & 0 & 0 & 0 & 1 & 1 & 1860297.32584311918651565065 \\ 
        1 & 0 & 0 & 0 & 0 & 2 & -8167.59106512353221907219 \\ 
        0 & 3 & 0 & 0 & 0 & 0 & 509.79487592937756404443 \\ 
        0 & 2 & 1 & 0 & 0 & 0 & -4371420.16210973240458770306 \\
        0 & 2 & 0 & 1 & 0 & 0 & 62.16310668137721201498 \\
        0 & 2 & 0 & 0 & 1 & 0 & -204549.45186648079749147655 \\
        0 & 2 & 0 & 0 & 0 & 1 & 253.43088282566487499936 \\ 
        0 & 1 & 2 & 0 & 0 & 0 & -15132334067.95542952418327331543 \\
        0 & 1 & 1 & 1 & 0 & 0 & -2217655.25460645515317992249 \\
        0 & 1 & 1 & 0 & 1 & 0 & 414176401.56812856081523932517 \\
        0 & 1 & 1 & 0 & 0 & 1 & -1180366.74579333393751312542 \\
        0 & 1 & 0 & 2 & 0 & 0 & -58.33007221422800885058 \\ 
        0 & 1 & 0 & 1 & 1 & 0 & 17828.63514419211671224730 \\
        0 & 1 & 0 & 1 & 0 & 1 & 51.70479214564455009220 \\ 
        0 & 1 & 0 & 0 & 2 & 0 & 13736.69772259018608195902 \\
        0 & 1 & 0 & 0 & 1 & 1 & 6830.43944911520275375949 \\
        0 & 1 & 0 & 0 & 0 & 2 & -20.67845605640393444151 \\
        0 & 0 & 3 & 0 & 0 & 0 & 381987821594144.40618896484375000000 \\
        0 & 0 & 2 & 1 & 0 & 0 & 2137975651.27025643410161137581 \\
        0 & 0 & 2 & 0 & 1 & 0 & -11786084090302.31474399566650390625 \\
        0 & 0 & 2 & 0 & 0 & 1 & 12830189977.95690678991377353668 \\
        0 & 0 & 1 & 2 & 0 & 0 & -3763590.07366334064204238530 \\
        0 & 0 & 1 & 1 & 1 & 0 & 169566375.25463584749377332628 \\
        0 & 0 & 1 & 1 & 0 & 1 & 50848.87687928509182100356 \\
        0 & 0 & 1 & 0 & 2 & 0 & 96637410.75323125508293742314 \\
        0 & 0 & 1 & 0 & 1 & 1 & 53395819.48057151796456309967 \\
        0 & 0 & 1 & 0 & 0 & 2 & -46935.92992026009458683689 \\
        0 & 0 & 0 & 3 & 0 & 0 & 385.83738973098191202271 \\
        0 & 0 & 0 & 2 & 1 & 0 & -90458.85125732885784799464 \\
        0 & 0 & 0 & 2 & 0 & 1 & -15.56798686589613507116 \\
        0 & 0 & 0 & 1 & 2 & 0 & 21541.55589805406764014606 \\
        0 & 0 & 0 & 1 & 1 & 1 & 4954.22010897005621021805 \\
        0 & 0 & 0 & 1 & 0 & 2 & -35.84774062125642758378 \\
    \end{tabular}
\end{table}
\begin{table}[H]
    \centering
    \begin{tabular}{l|l|l|l|l|l|l}
    \hline
        i & j & k & l & m & n & ML coefficient \\ \hline
        0 & 0 & 0 & 0 & 3 & 0 & -10294.67906127658141723913 \\
        0 & 0 & 0 & 0 & 2 & 1 & -1391.22492644369575054419 \\
        0 & 0 & 0 & 0 & 1 & 2 & -3506.49819360245294164180 \\
        0 & 0 & 0 & 0 & 0 & 3 & 3.22868098144071187885 \\
        4 & 0 & 0 & 0 & 0 & 0 & -609884206233062.21185302734375000000 \\
        3 & 1 & 0 & 0 & 0 & 0 & 2286649649761.54408359527587890625 \\
        3 & 0 & 1 & 0 & 0 & 0 & -11390421337335452.47851562500000000000 \\
        3 & 0 & 0 & 1 & 0 & 0 & 3108755170674.35331082344055175781 \\
        3 & 0 & 0 & 0 & 1 & 0 & 803652876494166.94177246093750000000 \\
        3 & 0 & 0 & 0 & 0 & 1 & -939182230289.88944846391677856445 \\
        2 & 2 & 0 & 0 & 0 & 0 & -1003944516.22098043363075703382 \\
        2 & 1 & 1 & 0 & 0 & 0 & 10411649205634.87721920013427734375 \\
        2 & 1 & 0 & 1 & 0 & 0 & -1557924806.51389433187432587147 \\
        2 & 1 & 0 & 0 & 1 & 0 & -118760568368.98034311085939407349 \\
        2 & 1 & 0 & 0 & 0 & 1 & 334652805.48934148746775463223 \\
        2 & 0 & 2 & 0 & 0 & 0 & -3317032918211318.67309570312500000000 \\
        2 & 0 & 1 & 1 & 0 & 0 & -9562724104511.23784637451171875000 \\
        2 & 0 & 1 & 0 & 1 & 0 & -384097323916005.53619384765625000000 \\
        2 & 0 & 1 & 0 & 0 & 1 & 2032748104464.98834764957427978516 \\
        2 & 0 & 0 & 2 & 0 & 0 & -1449139265.51888303633313626051 \\
        2 & 0 & 0 & 1 & 1 & 0 & 91697382186.32168483734130859375 \\
        2 & 0 & 0 & 1 & 0 & 1 & -11472230.71684562088921666145 \\
        2 & 0 & 0 & 0 & 2 & 0 & -26181694992.62494347244501113892 \\
        2 & 0 & 0 & 0 & 1 & 1 & 1978689273.68517350044567137957 \\
        2 & 0 & 0 & 0 & 0 & 2 & 33726104.11829262259925599210 \\
        1 & 3 & 0 & 0 & 0 & 0 & 216522.05698575852947840303 \\
        1 & 2 & 1 & 0 & 0 & 0 & 36508964.46534274275836651213 \\
        1 & 2 & 0 & 1 & 0 & 0 & 116058.91464810080243807988 \\
        1 & 2 & 0 & 0 & 1 & 0 & 2898539.99654126416021426849 \\
        1 & 2 & 0 & 0 & 0 & 1 & 25236.56844533908122407695 \\
        1 & 1 & 2 & 0 & 0 & 0 & -23073900158445.17253684997558593750 \\
        1 & 1 & 1 & 1 & 0 & 0 & 667469733.10537990240845829248 \\
        1 & 1 & 1 & 0 & 1 & 0 & -38834638418.39343107491731643677 \\
        1 & 1 & 1 & 0 & 0 & 1 & 220239798.15761969446612056345 \\
        1 & 1 & 0 & 2 & 0 & 0 & -51693.03549516992243795244 \\
        1 & 1 & 0 & 1 & 1 & 0 & -5828472.47446352284759996110 \\
        1 & 1 & 0 & 1 & 0 & 1 & 476.42540283899596861850 \\
        1 & 1 & 0 & 0 & 2 & 0 & 1147454.69245391907691100641 \\
        1 & 1 & 0 & 0 & 1 & 1 & 572959.34117592326998646968 \\
    \end{tabular}
\end{table}
\begin{table}[H]
    \centering
    \begin{tabular}{l|l|l|l|l|l|l}
    \hline
        i & j & k & l & m & n & ML coefficient \\ \hline
        1 & 1 & 0 & 0 & 0 & 2 & -4848.90596773981986045143 \\
        1 & 0 & 3 & 0 & 0 & 0 & 1185978908795400884.25000000000000000000 \\
        1 & 0 & 2 & 1 & 0 & 0 & 290262424530080.35983276367187500000 \\
        1 & 0 & 2 & 0 & 1 & 0 & 3132752484687146.03002929687500000000 \\
        1 & 0 & 2 & 0 & 0 & 1 & -23748440602989.03187370300292968750 \\
        1 & 0 & 1 & 2 & 0 & 0 & -4403567830.39056837186217308044 \\
        1 & 0 & 1 & 1 & 1 & 0 & 34305792027.43660329654812812805 \\
        1 & 0 & 1 & 1 & 0 & 1 & -203729258.44310540329024661332 \\
        1 & 0 & 1 & 0 & 2 & 0 & -41186722047.42009187862277030945 \\
        1 & 0 & 1 & 0 & 1 & 1 & -98070603073.36613352596759796143 \\
        1 & 0 & 1 & 0 & 0 & 2 & 101467077.37953558584558777511 \\
        1 & 0 & 0 & 3 & 0 & 0 & 922536.43432689239870114761 \\
        1 & 0 & 0 & 2 & 1 & 0 & 28427085.05554292692067974713 \\
        1 & 0 & 0 & 2 & 0 & 1 & 78619.04986536038313005292 \\
        1 & 0 & 0 & 1 & 2 & 0 & 1053203.41301598106258552434 \\
        1 & 0 & 0 & 1 & 1 & 1 & 289387.22330324230449605238 \\
        1 & 0 & 0 & 1 & 0 & 2 & -2694.99654954760472369557 \\
        1 & 0 & 0 & 0 & 3 & 0 & 1372063.51610930279696276557 \\
        1 & 0 & 0 & 0 & 2 & 1 & 512290.41829282706618187149 \\
        1 & 0 & 0 & 0 & 1 & 2 & 111095.92372743804822476932 \\
        1 & 0 & 0 & 0 & 0 & 3 & -767.34723686879926507487 \\
        0 & 4 & 0 & 0 & 0 & 0 & -163.88545967769210442988 \\
        0 & 3 & 1 & 0 & 0 & 0 & 2666212.68596806765071960399 \\
        0 & 3 & 0 & 1 & 0 & 0 & 206.89413650069255329322 \\
        0 & 3 & 0 & 0 & 1 & 0 & 135209.91901166244183229992 \\
        0 & 3 & 0 & 0 & 0 & 1 & -214.31191652850269152575 \\
        0 & 2 & 2 & 0 & 0 & 0 & -9136732506.98263535927981138229 \\
        0 & 2 & 1 & 1 & 0 & 0 & -279379.51584819914089052872 \\
        0 & 2 & 1 & 0 & 1 & 0 & -66616029.03715856408962281421 \\
        0 & 2 & 1 & 0 & 0 & 1 & 180913.44472203148671951567 \\
        0 & 2 & 0 & 2 & 0 & 0 & -419.57531205216068040387 \\
        0 & 2 & 0 & 1 & 1 & 0 & 1489.14045495658500473102 \\
        0 & 2 & 0 & 1 & 0 & 1 & 11.23875303290998837937 \\
        0 & 2 & 0 & 0 & 2 & 0 & -1358.59014182990396679696 \\
        0 & 2 & 0 & 0 & 1 & 1 & -1390.48499983711566097000 \\
        0 & 2 & 0 & 0 & 0 & 2 & 9.87851840463800074093 \\
        0 & 1 & 3 & 0 & 0 & 0 & 462918671706828.88278198242187500000 \\
        0 & 1 & 2 & 1 & 0 & 0 & -15241873242.83994901087135076523 \\
        0 & 1 & 2 & 0 & 1 & 0 & -1312372475306.72573256492614746094 \\
    \end{tabular}
\end{table}
\begin{table}[H]
    \centering
    \begin{tabular}{l|l|l|l|l|l|l}
    \hline
        i & j & k & l & m & n & ML coefficient \\ \hline
        0 & 1 & 2 & 0 & 0 & 1 & 1104286645.31508435343857854605 \\
        0 & 1 & 1 & 2 & 0 & 0 & 2059660.58497703978571280459 \\
        0 & 1 & 1 & 1 & 1 & 0 & -51345925.33649688777222763747 \\
        0 & 1 & 1 & 1 & 0 & 1 & 297448.77424125082683303845 \\
        0 & 1 & 1 & 0 & 2 & 0 & -47377338.73872763563485932536 \\
        0 & 1 & 1 & 0 & 1 & 1 & -24839992.79171100644634861965 \\
        0 & 1 & 1 & 0 & 0 & 2 & 107436.55368837489916700179 \\
        0 & 1 & 0 & 3 & 0 & 0 & 670.41295937613563904467 \\
        0 & 1 & 0 & 2 & 1 & 0 & -10821.33304008068227552997 \\
        0 & 1 & 0 & 2 & 0 & 1 & -79.87132008716167705376 \\
        0 & 1 & 0 & 1 & 2 & 0 & -879.36891396402831727608 \\
        0 & 1 & 0 & 1 & 1 & 1 & -809.86494124568962010846 \\
        0 & 1 & 0 & 1 & 0 & 2 & 1.82199112992164052416 \\
        0 & 1 & 0 & 0 & 3 & 0 & -1420.27811165072932109155 \\
        0 & 1 & 0 & 0 & 2 & 1 & -836.31408028778357899302 \\
        0 & 1 & 0 & 0 & 1 & 2 & -351.27098533628643167726 \\
        0 & 1 & 0 & 0 & 0 & 3 & 0.71905927455454144820 \\
        0 & 0 & 4 & 0 & 0 & 0 & -1798640501668048071.5 \\
        0 & 0 & 3 & 1 & 0 & 0 & -262912583188294.60734558105468750000 \\
        0 & 0 & 3 & 0 & 1 & 0 & 55397461333202594.48046875000000000000 \\
        0 & 0 & 3 & 0 & 0 & 1 & -98395064336925.39571380615234375000 \\
        0 & 0 & 2 & 2 & 0 & 0 & 16536854822.98794694989919662476 \\
        0 & 0 & 2 & 1 & 1 & 0 & 1365382879206.47664892673492431641 \\
        0 & 0 & 2 & 1 & 0 & 1 & 47567040.79264765448169782758 \\
        0 & 0 & 2 & 0 & 2 & 0 & 1774753452474.46298027038574218750 \\
        0 & 0 & 2 & 0 & 1 & 1 & 474303489991.20203351974487304688 \\
        0 & 0 & 2 & 0 & 0 & 2 & -822571369.61729244329035282135 \\
        0 & 0 & 1 & 3 & 0 & 0 & 4119090.52285701097980563645 \\
        0 & 0 & 1 & 2 & 1 & 0 & -118399882.27104566253547091037 \\
        0 & 0 & 1 & 2 & 0 & 1 & -215297.47447441806636447836 \\
        0 & 0 & 1 & 1 & 2 & 0 & -21838955.79877393666902207769 \\
        0 & 0 & 1 & 1 & 1 & 1 & -7276476.24265170020180448773 \\
        0 & 0 & 1 & 1 & 0 & 2 & 15949.43295318960734885394 \\
        0 & 0 & 1 & 0 & 3 & 0 & -16945880.44705596407766279299 \\
        0 & 0 & 1 & 0 & 2 & 1 & -14052848.57842917150355788181 \\
        0 & 0 & 1 & 0 & 1 & 2 & -5140361.12635335735694752657 \\
        0 & 0 & 1 & 0 & 0 & 3 & 3747.54514574331018117448 \\
        0 & 0 & 0 & 4 & 0 & 0 & 243.50975039281769550514 \\
        0 & 0 & 0 & 3 & 1 & 0 & 139610.95607158872688557949 \\
    \end{tabular}
\end{table}
\begin{table}[H]
    \centering
    \begin{tabular}{l|l|l|l|l|l|l}
    \hline
        i & j & k & l & m & n & ML coefficient \\ \hline
        0 & 0 & 0 & 3 & 0 & 1 & -239.55997636072933952567 \\
        0 & 0 & 0 & 2 & 2 & 0 & -15469.59544613538603829994 \\
        0 & 0 & 0 & 2 & 1 & 1 & -2392.33852405087376613579 \\
        0 & 0 & 0 & 2 & 0 & 2 & 38.70380969332508522621 \\
        0 & 0 & 0 & 1 & 3 & 0 & -653.24534710175902557650 \\
        0 & 0 & 0 & 1 & 2 & 1 & -338.59715065214459644993 \\
        0 & 0 & 0 & 1 & 1 & 2 & -162.89119741277407200919 \\
        0 & 0 & 0 & 1 & 0 & 3 & -0.37540207904611577557 \\
        0 & 0 & 0 & 0 & 4 & 0 & 1399.17742189930837970469 \\
        0 & 0 & 0 & 0 & 3 & 1 & 821.90471373131219229302 \\
        0 & 0 & 0 & 0 & 2 & 2 & 80.07987857983952287300 \\
        0 & 0 & 0 & 0 & 1 & 3 & 293.04664325126301419044 \\
        0 & 0 & 0 & 0 & 0 & 4 & -0.20554091389092022358 \\
    \end{tabular}
\end{table}
\subsection{Appendix E Lattice scales}
\begin{longtable}{c|c|c|c|c|c}
\caption{Dataset} \\
\hline
Col1 & Col2 & Col3 & Col4 & Col5 & Col6 \\
\hline
\endfirsthead

\hline
Col1 & Col2 & Col3 & Col4 & Col5 & Col6 \\
\hline
\endhead

1 & 10 & 1 & 1 & 1 & 24443.1 \\
2 & 10 & 1 & 1 & 0.1 & 24830.7 \\
3 & 10 & 1 & 1 & 0.01 & 25125.4 \\
4 & 10 & 1 & 1 & 0 & 24638.5 \\
5 & 10 & 1 & 0.6 & 1 & 24130 \\
6 & 10 & 1 & 0.6 & 0.1 & 23153.8 \\
7 & 10 & 1 & 0.6 & 0.01 & 23633.7 \\
8 & 10 & 1 & 0.6 & 0 & 25536.7 \\
9 & 10 & 1 & 0.001 & 1 & 23466.4 \\
10 & 10 & 1 & 0.001 & 0.1 & 23290.2 \\
11 & 10 & 1 & 0.001 & 0.01 & 24041.7 \\
12 & 10 & 1 & 0.001 & 0 & 23675 \\
13 & 10 & 0.6 & 1 & 1 & 24387.5 \\
14 & 10 & 0.6 & 1 & 0.1 & 25544.9 \\
15 & 10 & 0.6 & 1 & 0.01 & 23425.4 \\
16 & 10 & 0.6 & 1 & 0 & 23949.2 \\
17 & 10 & 0.6 & 0.6 & 1 & 24450.7 \\
18 & 10 & 0.6 & 0.6 & 0.1 & 22963 \\
19 & 10 & 0.6 & 0.6 & 0.01 & 23265.1 \\
20 & 10 & 0.6 & 0.6 & 0 & 23889.4 \\
21 & 10 & 0.6 & 0.001 & 1 & 24849.4 \\
22 & 10 & 0.6 & 0.001 & 0.1 & 24682.3 \\
23 & 10 & 0.6 & 0.001 & 0.01 & 24456.8 \\
24 & 10 & 0.6 & 0.001 & 0 & 24255 \\
25 & 10 & $10^{-6}$ & 1 & 1 & 22973.4 \\
26 & 10 & $10^{-6}$ & 1 & 0.1 & 24416.3 \\
27 & 10 & $10^{-6}$ & 1 & 0.01 & 23143.9 \\
28 & 10 & $10^{-6}$ & 1 & 0 & 27009.6 \\
29 & 10 & $10^{-6}$ & 0.6 & 1 & 24377.6 \\
30 & 10 & $10^{-6}$ & 0.6 & 0.1 & 25216.5 \\
31 & 10 & $10^{-6}$ & 0.6 & 0.01 & 28112.6 \\
32 & 10 & $10^{-6}$ & 0.6 & 0 & 24178.3 \\
33 & 10 & $10^{-6}$ & 0.001 & 1 & 23213.8 \\
34 & 10 & $10^{-6}$ & 0.001 & 0.1 & 22978.2 \\
35 & 10 & $10^{-6}$ & 0.001 & 0.01 & 25976.7 \\
36 & 10 & $10^{-6}$ & 0.001 & 0 & 25044.6 \\
37 & 10 & $10^{-12}$ & 1 & 1 & $2.00372\times10^{9}$ \\
38 & 10 & $10^{-12}$ & 1 & 0.1 & $1.04766\times10^{10}$ \\
39 & 10 & $10^{-12}$ & 1 & 0.01 & $9.80958\times10^{7}$ \\
40 & 10 & $10^{-12}$ & 1 & 0 & $2.06461\times10^{8}$ \\

41 & 10 & $10^{-12}$ & 0.6 & 1 & $2.2722\times10^{6}$ \\
42 & 10 & $10^{-12}$ & 0.6 & 0.1 & $7.19458\times10^{8}$ \\
43 & 10 & $10^{-12}$ & 0.6 & 0.01 & $3.33201\times10^{9}$ \\
44 & 10 & $10^{-12}$ & 0.6 & 0 & $2.49569\times10^{9}$ \\
45 & 10 & $10^{-12}$ & 0.001 & 1 & $2.03993\times10^{8}$ \\
46 & 10 & $10^{-12}$ & 0.001 & 0.1 & $1.46879\times10^{9}$ \\
47 & 10 & $10^{-12}$ & 0.001 & 0.01 & $1.98615\times10^{9}$ \\
48 & 10 & $10^{-12}$ & 0.001 & 0 & $2.41196\times10^{9}$ \\
49 & 1 & 1 & 1 & 1 & $1.59809\times10^{9}$ \\
50 & 1 & 1 & 1 & 0.1 & $2.39939\times10^{8}$ \\
51 & 1 & 1 & 1 & 0.01 & $1.70595\times10^{10}$ \\
52 & 1 & 1 & 1 & 0 & $9.04416\times10^{8}$ \\
53 & 1 & 1 & 0.6 & 1 & $4.8067\times10^{6}$ \\
54 & 1 & 1 & 0.6 & 0.1 & $2.96235\times10^{10}$ \\
55 & 1 & 1 & 0.6 & 0.01 & $1.17841\times10^{10}$ \\
56 & 1 & 1 & 0.6 & 0 & $1.3323\times10^{9}$ \\
57 & 1 & 1 & 0.001 & 1 & $1.44283\times10^{6}$ \\
58 & 1 & 1 & 0.001 & 0.1 & $2.349\times10^{9}$ \\
59 & 1 & 1 & 0.001 & 0.01 & $8.06523\times10^{8}$ \\
60 & 1 & 1 & 0.001 & 0 & $2.54969\times10^{10}$ \\

61 & 1 & 0.6 & 1 & 1 & $1.72839\times10^{10}$ \\
62 & 1 & 0.6 & 1 & 0.1 & $1.36135\times10^{8}$ \\
63 & 1 & 0.6 & 1 & 0.01 & $2.00186\times10^{10}$ \\
64 & 1 & 0.6 & 1 & 0 & $8.94893\times10^{8}$ \\
65 & 1 & 0.6 & 0.6 & 1 & $4.75048\times10^{8}$ \\
66 & 1 & 0.6 & 0.6 & 0.1 & $1.06328\times10^{9}$ \\
67 & 1 & 0.6 & 0.6 & 0.01 & $1.72761\times10^{9}$ \\
68 & 1 & 0.6 & 0.6 & 0 & $1.05257\times10^{9}$ \\
69 & 1 & 0.6 & 0.001 & 1 & $5.54004\times10^{9}$ \\
70 & 1 & 0.6 & 0.001 & 0.1 & 599070 \\
71 & 1 & 0.6 & 0.001 & 0.01 & $2.34235\times10^{9}$ \\
72 & 1 & 0.6 & 0.001 & 0 & $5.11756\times10^{8}$ \\
73 & 1 & $10^{-6}$ & 1 & 1 & $6.27217\times10^{9}$ \\
74 & 1 & $10^{-6}$ & 1 & 0.1 & $1.03525\times10^{9}$ \\
75 & 1 & $10^{-6}$ & 1 & 0.01 & $3.32658\times10^{6}$ \\
76 & 1 & $10^{-6}$ & 1 & 0 & $3.38809\times10^{6}$ \\
77 & 1 & $10^{-6}$ & 0.6 & 1 & $3.87222\times10^{6}$ \\
78 & 1 & $10^{-6}$ & 0.6 & 0.1 & $1.78477\times10^{9}$ \\
79 & 1 & $10^{-6}$ & 0.6 & 0.01 & 23033.5 \\
80 & 1 & $10^{-6}$ & 0.6 & 0 & $1.36746\times10^{10}$ \\
81 & 1 & $10^{-6}$ & 0.001 & 1 & $4.95376\times10^{9}$ \\
82 & 1 & $10^{-6}$ & 0.001 & 0.1 & $1.35714\times10^{8}$ \\
83 & 1 & $10^{-6}$ & 0.001 & 0.01 & $3.45213\times10^{6}$ \\
84 & 1 & $10^{-6}$ & 0.001 & 0 & $5.1159\times10^{9}$ \\
85 & 1 & $10^{-12}$ & 1 & 1 & $6.27217\times10^{9}$ \\
86 & 1 & $10^{-12}$ & 1 & 0.1 & $1.03525\times10^{9}$ \\
87 & 1 & $10^{-12}$ & 1 & 0.01 & $3.32658\times10^{6}$ \\
88 & 1 & $10^{-12}$ & 1 & 0 & $3.38809\times10^{6}$ \\
89 & 1 & $10^{-12}$ & 0.6 & 1 & $3.87222\times10^{6}$ \\
90 & 1 & $10^{-12}$ & 0.6 & 0.1 & $1.78477\times10^{9}$ \\
91 & 1 & $10^{-12}$ & 0.6 & 0.01 & $4.93188\times10^{8}$ \\
92 & 1 & $10^{-12}$ & 0.6 & 0 & $1.36746\times10^{10}$ \\
93 & 1 & $10^{-12}$ & 0.001 & 1 & $4.95376\times10^{9}$ \\
94 & 1 & $10^{-12}$ & 0.001 & 0.1 & $1.35714\times10^{8}$ \\
95 & 1 & $10^{-12}$ & 0.001 & 0.01 & $3.45213\times10^{6}$ \\
96 & 1 & $10^{-12}$ & 0.001 & 0 & $5.1159\times10^{9}$ \\
97 & 0.001 & 1 & 1 & 1 & 46068.9 \\
98 & 0.001 & 1 & 1 & 0.1 & 44760.6 \\
99 & 0.001 & 1 & 1 & 0.01 & 45795.2 \\
100 & 0.001 & 1 & 1 & 0 & 45492.1 \\
101 & 0.001 & 1 & 0.6 & 1 & 49014 \\
102 & 0.001 & 1 & 0.6 & 0.1 & 52368.7 \\
103 & 0.001 & 1 & 0.6 & 0.01 & 62566.2 \\
104 & 0.001 & 1 & 0.6 & 0 & 60472 \\
105 & 0.001 & 1 & 0.001 & 1 & 47172.3 \\
106 & 0.001 & 1 & 0.001 & 0.1 & 546525 \\
107 & 0.001 & 1 & 0.001 & 0.01 & 54029.4 \\
108 & 0.001 & 1 & 0.001 & 0 & 38540.7 \\
109 & 0.001 & 0.6 & 1 & 1 & 72798.3 \\
110 & 0.001 & 0.6 & 1 & 0.1 & 62217.2 \\
111 & 0.001 & 0.6 & 1 & 0.01 & 47483 \\
112 & 0.001 & 0.6 & 1 & 0 & 47352.3 \\
113 & 0.001 & 0.6 & 0.6 & 1 & 37276.8 \\
114 & 0.001 & 0.6 & 0.6 & 0.1 & 73326.2 \\
115 & 0.001 & 0.6 & 0.6 & 0.01 & 38063.9 \\
116 & 0.001 & 0.6 & 0.6 & 0 & 44735.5 \\
117 & 0.001 & 0.6 & 0.001 & 1 & 44495.1 \\
118 & 0.001 & 0.6 & 0.001 & 0.1 & 45495.2 \\
119 & 0.001 & 0.6 & 0.001 & 0.01 & 49017.3 \\
120 & 0.001 & 0.6 & 0.001 & 0 & 45611.9 \\
121 & 0.001 &  1e-06 & 1 & 1 & 39450.7 \\
122 &  0.001 &  1e-06 & 1 & 0.1 & 9.10176e+07 \\
123 &  0.001 &  1e-06 & 1 & 0.01 & 46738.9 \\
124 &  0.001 &  1e-06 &  1 & 0 & 73997.9 \\
125 &  0.001 &  1e-06 &  0.6 & 1 & 50283.8 \\
126 &  0.001 &  1e-06 &  0.6 & 0.1 & 34969.4 \\
127 &  0.001 &  1e-06 &  0.6 & 0.01 & 46868.7 \\
128 &  0.001 &  1e-06 &  0.6 & 0 & 56439.4 \\
129 &  0.001 &  1e-06 &  0.001 & 1 & 40090.2 \\
130 &  0.001 &  1e-06 &  0.001 & 0.1 & 46910.9 \\
131 &  0.001 &  1e-06 &  0.001 & 0.01 & 48004.1 \\
132 &  0.001 &  1e-06 &  0.001 & 0 & 41773.3 \\
133 &  0.001 &  1e-12 &  1 & 1 & 39450.7 \\
134 &  0.001 &  1e-12 &  1 & 0.1 & 9.10176e+07 \\
135 &  0.001 &  1e-12 &  1 & 0.01 & 46738.9 \\
136 &  0.001 &  1e-12 &  1 & 0 & 73997.9 \\
137 &  0.001 &  1e-12 &  0.6 & 1 & 50283.8 \\
138 &  0.001 &  1e-12 &  0.6 & 0.1 & 34969.4 \\
139 &  0.001 &  1e-12 &  0.6 & 0.01 & 46868.7 \\
140 &  0.001 &  1e-12 &  0.6 & 0 & 56439.4 \\
141 &  0.001 &  1e-12 &  0.001 & 1 & 40090.2 \\
142 &  0.001 &  1e-12 &  0.001 & 0.1 & 46910.9 \\
143 &  0.001 &  1e-12 &  0.001 & 0.01 & 48004.1 \\
144 &  0.001 &  1e-12 &  0.001 & 0 & 49849.2 \\
145 &  1e-06 &  1 & 1 &  1 & 2.21865e+08 \\
146 &  1e-06 &  1 & 1 &  0.1 & 6.11409e+09 \\
147 &  1e-06 &  1 &  1 & 0.01 & 3.24112e+10 \\
148 &  1e-06 &  1 &  1 & 0 & 2.45424e+09 \\
149 &  1e-06 &  1 &  0.6 & 1 & 2.49933e+08 \\
150 &  1e-06 &  1 &  0.6 & 0.1 & 1.59103e+10 \\
151 &  1e-06 &  1 &  0.6 & 0.01 & 1.35976e+09 \\
152 &  1e-06 &  1 &  0.6 & 0 & 9.52742e+07 \\
153 &  1e-06 &  1 &  0.001 & 1 & 9.67767e+08 \\
154 &  1e-06 &  1 &  0.001 & 0.1 & 1.43128e+08 \\
155 &  1e-06 &  1 &  0.001 & 0.01 & 1.00044e+10 \\
156 &  1e-06 &  1 &  0.001 & 0 & 7.75744e+08 \\
157 &  1e-06 &  0.6 & 1 & 1 & 6.96716e+08 \\
158 &  1e-06 &  0.6 & 1 & 0.1 & 2.09134e+09 \\
159 &  1e-06 &  0.6 &  1 & 0.01 & 1.21266e+09 \\
160 &  1e-06 &  0.6 &  1 & 0 & 1.29298e+10 \\
161 &  1e-06 &  0.6 &  0.6 & 1 & 1.18021e+09 \\
162 &  1e-06 &  0.6 &  0.6 & 0.1 & 5.84978e+10 \\
163 &  1e-06 &  0.6 &  0.6 & 0.01 & 2.53722e+08 \\
164 &  1e-06 &  0.6 &  0.6 & 0 & 1.89108e+09 \\
165 &  1e-06 &  0.6 & 0.001 & 1 & 1.91593e+10 \\
166 &  1e-06 &  0.6 &  0.001 & 0.1 & 9.84041e+09 \\
167 &  1e-06 &  0.6 &  0.001 & 0.01 & 4.08262e+09 \\
168 &  1e-06 &  0.6 &  0.001 & 0 & 9.61274e+08 \\
169 &  1e-06 &  1e-06 & 1 & 1 & 1.61208e+09 \\
170 &  1e-06 &  1e-06 & 1 & 0.1 & 9.35359e+09 \\
171 &  1e-06 &  1e-06 & 1 & 0.01 & 9.86342e+09 \\
172 &  1e-06 &  1e-06 & 1 & 0 & 6.057e+08 \\
173 &  1e-06 &  1e-06 & 0.6 & 1 & 2.4442e+09 \\
174 &  1e-06 &  1e-06 & 0.6 & 0.1 & 2.00464e+09 \\
175 &  1e-06 &  1e-06 & 0.6 & 0.01 & 8.42583e+08 \\
176 &  1e-06 &  1e-06 & 0.6 & 0 & 1.72654e+09 \\
177 &  1e-06 &  1e-06 & 0.001 & 1 & 1.82719e+10 \\
178 &  1e-06 &  1e-06 & 0.001 & 0.1 & 3.32334e+09 \\
179 &  1e-06 &  1e-06 & 0.001 & 0.01 & 2.84855e+09 \\
180 &  1e-06 &  1e-06 & 0.001 & 0 & 9.49918e+09 \\
181 &  1e-06 &  1e-12 & 1 & 1 & 1.61208e+09 \\
182 &  1e-06 &  1e-12 & 1 & 0.1 & 9.35359e+09 \\
183 &  1e-06 &  1e-12 & 1 & 0.01 & 9.86342e+09 \\
184 &  1e-06 &  1e-12 & 1 & 0 & 6.057e+08 \\
185 &  1e-06 &  1e-12 & 0.6 & 1 & 2.4442e+09 \\
186 &  1e-06 &  1e-12 & 0.6 & 0.1 & 2.00464e+09 \\
187 &  1e-06 &  1e-12 & 0.6 & 0.01 & 8.42583e+08 \\
188 &  1e-06 &  1e-12 & 0.6 & 0 & 1.72654e+09 \\
189 &  1e-06 &  1e-12 & 0.001 & 1 & 1.82719e+10 \\
190 &  1e-06 &  1e-12 & 0.001 & 0.1 & 3.32334e+09 \\
191 &  1e-06 &  1e-12 & 0.001 & 0.01 & 2.84855e+09 \\
192 &  1e-06 &  1e-12 & 0.001 & 0 & 9.49918e+09 \\
\end{longtable}
\bibliographystyle{plain}
\bibliography{bib}
\end{document}